\documentclass[a4paper,11pt]{article}
\pdfoutput=1 
\usepackage{jcappub} 

\usepackage{mathalfa}
\usepackage{graphicx}
\usepackage{caption}
\usepackage{subcaption}
\usepackage{amsmath,amssymb}
\usepackage[font=small,labelfont=bf]{caption}
\usepackage{verbatim}
\usepackage{placeins}
\usepackage{tikz}
\usetikzlibrary{arrows, positioning, angles,quotes, calc, intersections}
\usetikzlibrary{decorations.pathreplacing}
\def\mean#1{\left< #1 \right>}

\usepackage{color}

\usepackage[normalem]{ulem}

\title{Title}

\newcommand{\ndv}{{v_{\parallel}}}

\newcommand{\bk}{{\mathbf k}}

\newcommand{\bn}{{\mathbf n}}
\newcommand{\bv}{{\mathbf v}}

\newcommand{\be}{\begin{equation}}
\newcommand{\ee}{\end{equation}}

\newcommand{\lsim}{\stackrel{<}{\sim}}
\newcommand{\bea}{\begin{eqnarray}}
\newcommand{\eea}{\end{eqnarray}}
\newcommand{\bean}{\begin{eqnarray*}}
\newcommand{\eean}{\end{eqnarray*}}

\newcommand{\HH}{{\cal H}}

\author[a,b,c]{Francesca~Lepori,}
\author[d,e,f]{Vid~Ir\v{s}i\v{c},}
\author[g,h,i]{Enea~Di~Dio,}
\author[b,c,j,k]{Matteo~Viel}

\affiliation[a]{Department de Physique Theorique and Center for Astroparticle Physics, Universite de Geneve, Quai E. Ansermet 24, CH-1211 Geneve 4, Switzerland}
\affiliation[b]{SISSA- International School for Advanced Studies, Via Bonomea 265, 34136 Trieste, Italy}
\affiliation[c]{INFN, Sezione di Trieste, Via Valerio 2, I-34127 Trieste, Italy}
\affiliation[d]{Kavli Institute for Cosmology, University of Cambridge, Madingley Road, Cambridge CB3 0HA, UK}
\affiliation[e]{Cavendish Laboratory, University of Cambridge, 19 J. J. Thomson Ave., Cambridge CB3 0HE, UK}
\affiliation[f]{University of Washington, Department of Astronomy, 3910 15th Ave NE, WA 98195-1580 Seattle, USA}
\affiliation[g]{Lawrence Berkeley National Laboratory, 1 Cyclotron Road, Berkeley, CA 93720, USA}
\affiliation[h]{Berkeley Center for Cosmological Physics and Department of Physics, University of California, Berkeley, CA 94720}
\affiliation[i]{Center for Theoretical Astrophysics and Cosmology, Institute for Computational Science, University of Zurich, Winterthurerstrasse 190, CH-8057 Zurich, Switzerland}
\affiliation[j]{INAF - Osservatorio Astronomico di Trieste,  Via G. B. Tiepolo 11,  I-34143 Trieste, Italy}
\affiliation[k]{IFPU, Institute for Fundamental Physics of the Universe, via Beirut 2, 34151 Trieste, Italy}

\emailAdd{Francesca.Lepori@unige.ch}
\emailAdd{vi223@cam.ac.uk}
\emailAdd{enea.didio@uzh.ch}
\emailAdd{viel@sissa.it}

\abstract{
We study the impact of relativistic effects in the 3-dimensional cross-correlation between Lyman-$\alpha$ forest and quasars. Apart from the relativistic effects, which are dominated by the Doppler contribution, several systematic effects are also included in our analysis (intervening metals, unidentified high column density systems, transverse proximity effect and effect of the UV fluctuations). We compute the signal-to-noise ratio for the Baryonic Oscillation Spectroscopic Survey (BOSS), the extended Baryonic Oscillation Spectroscopic Survey (eBOSS) and the Dark Energy Spectroscopic Instrument (DESI) surveys, showing that DESI will be able to detect the Doppler contribution in a Large Scale Structure (LSS) survey for the first time, with a S/N $>7$ for $r_{\rm min} > 10$ Mpc$/h$, where r$_{\rm min}$ denotes the minimum comoving separation between sources. We demonstrate that several physical parameters, introduced to provide a full modelling of the cross-correlation function, are affected by the Doppler contribution. By using a Fisher matrix approach, we establish that if the Doppler contribution is neglected in the data analysis, the derived parameters will be shifted by a non-negligible amount for the upcoming surveys.}

\begin{document}
\title{The impact of relativistic effects on the 3D Quasar-Lyman-$\alpha$ cross-correlation}
\maketitle
\flushbottom

\section{Introduction}

The spectra of high redshift quasars (QSOs) present a series of redshifted absorption lines due to Lyman-$\alpha$ (Ly$\alpha$) absorption from intervening neutral hydrogen in the intergalactic medium (IGM). These features are denoted as Lyman-$\alpha$ forest. 
In the past decade, the Ly$\alpha$ forest has been proven to be a rich source of cosmological information as tracer of the large scale structure \cite{Meiksin:2007rz} from large to small scales in a unique range of redshifts $z=[2-5.5]$.

In the past few years, the baryon acoustic feature has been detected by the survey SDSS-III/BOSS (Sloan Digital Sky Survey-III/Baryon Oscillation Spectroscopic Survey), both in the Ly$\alpha$
auto-correlation \cite{Busca2013, Slosar:2013fi, Delubac:2014aqe, Bautista:2017zgn} and in the cross-correlation
between Ly$\alpha$ and QSOs \cite{Font-Ribera:2013wce, Bourboux:2017cbm}.
A more recent analysis has been also presented 
with the first eBOSS (extended Baryon Oscillation Spectroscopic Survey) data \cite{Blomqvist:2019rah}, which improved the previous BOSS analysis. 

These analyses are based on the measurements of the 3D correlation function of the
transmitted forest flux. In order to fully exploit the cosmological information encoded in the data,
it is thus of crucial importance to include all the relevant physical effects in the modelling of the flux correlations. 
This will be an essential point for the next generation of surveys like DESI~\cite{desi0} and WEAVE-QSO~\cite{Pieri:2016jwo}, which are expected to significantly improve over the BOSS and eBOSS results.

In this work we focus on the so-called relativistic
effects, i.e.~the corrections to the Newtonian approximation due to the projection effects along the observed past light-cone, accounting for the light propagation in a clumpy universe form the source to the observer~\cite{yooA, yooB, bonvin_durrer, challinor_lewis,Schmidt:2012ne}.
In fact, it has been pointed out in the literature that some of these effects, e.g.~the Doppler correction, source the imaginary part
of the Fourier space power spectrum or, equivalently, the anti-symmetric part of the correlation function when
tracers with different biases are cross-correlated
\cite{mcdonald:2009, Bonvin:2013}.

In Ref.~\cite{Irsic:2015}, relativistic effects have been investigated in 
the cross-correlation between the forest transmitted flux and the quasars number counts fluctuations. This combination of tracers
is a promising target for the detection of such effects. Indeed, the signal is proportional to the bias difference between the two tracers, which
is large for this cross-correlation. 
In Ref.~\cite{Irsic:2015} the fully relativistic 
correlation function Ly$\alpha$-QSOs has been computed in linear theory and its several contributions have been quantified for the radial
correlation function. 
A signal-to-noise analysis for the imaginary part of Fourier space spectrum has also been presented
and it was shown that DESI should be able to 
detect this signal. 

The aim of this work is to extend the 
analysis presented in Ref.~\cite{Irsic:2015} to
the full 3D correlation function and with a more realistic and physical model, which includes all the astrophysical contaminants currently addressed by BOSS analysis~\cite{Bourboux:2017cbm}.
We finally intend to assess whether relativistic effects might constitute an important systematic for future Ly$\alpha$-QSOs analysis and
whether their detection is plausible in the near future. A first attempt to 
detect the relativistic effects in the Ly$\alpha$-QSOs correlations has been made recently with eBOSS data
\cite{Blomqvist:2019rah}. In this analysis, the relativistic corrections have been modelled as a dipole contribution to the cross-correlation 
and its amplitude treated as a free parameter to be fitted. The best-fit value points towards a non-zero amplitude for the dipole. However, since the amplitude is degenerate with other parameters, a clear detection of the effect could not be claimed.

This is the outline of this paper: in section \ref{boss-mod} we describe the BOSS model for the 3D correlation function, in section \ref{sec:rel-Eff} we report the fully relativistic
expression for the transmitted flux fluctuation and the quasars number count, in section \ref{sec:rel-Eff-2} we discuss the effects that 
may be relevant for our analysis and how they 
contribute to the 3D cross-correlation, in section \ref{sec:sn} we show the results of our signal-to-noise analysis and discuss whether we expect to detect the relativistic effects with DESI, in section \ref{sec:fisher} we study the impact of neglecting the relativistic effects 
in DESI on the parameter estimation, in section \ref{sec:conc} we summarize our main results and draw the conclusions. 

In the remainder of the paper, we will assume
a flat-$\Lambda$CDM cosmology. 
The fiducial values of the cosmological parameters 
are fixed to the values reported in Ref.~\cite{Bourboux:2017cbm} for the data analysis, Table 1.

\section{BOSS model for the Ly$\alpha$-quasar cross-correlation function}
\label{boss-mod}
In Ref.~\cite{Bourboux:2017cbm} the BOSS collaboration adopts a theoretical model for the observed 
cross-correlation which includes contamination from metals, High Column Density systems (HCDs), proximity effect and the effect of the UV fluctuations on the Ly$\alpha$ large scale clustering.
Here we summarize the features of this model, which will be our reference along the manuscript.
The observed correlation function is a function of the observed coordinates $(z_1,z_2, \cos\theta = \bn_1 \cdot\bn_2)$, which denotes the redshifts and the angle between the two directions\footnote{We use the convention adopted in Refs~\cite{bonvin_durrer,CLASSgal}, where $\bn$ indicates the direction from the source to the observer. For sake of simplicity we refer to $\bn$ as the direction on the sky, that formally is $-\bn$.} in the sky of the two sources. By assuming a cosmological model we can convert the observed angles and redshifts into distances by introducing a more convenient coordinate system $\left(z, r_\perp, r_\parallel \right)$ defined through
\begin{equation}
r_\perp = (D_\alpha + D_\text{Q}) \sin{\biggl(\frac{\theta}{2}\biggr)}, \qquad r_\parallel = (D_\alpha - D_\text{Q}) \cos{\biggl(\frac{\theta}{2}\biggr)},
\end{equation}
where $D_\alpha$ and $D_\text{Q}$ are the comoving distances to the Ly$\alpha$ and QSOs positions, respectively.

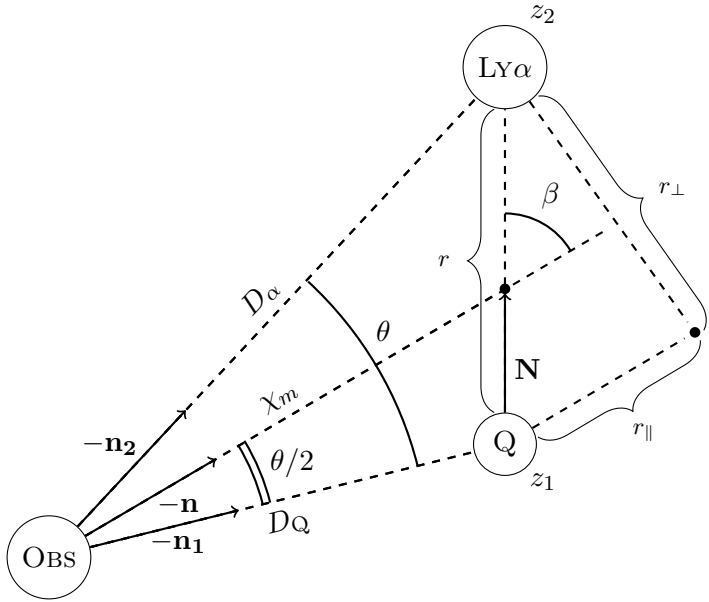
\begin{figure}
\begin{center}
\begin{tikzpicture}[scale=1.0] 
\node[draw,circle] (B) at (0,-2.5) {\textsc{Obs}};
\node[] (A1) at (1.7,-2.3) {$-\mathbf{n_1}$};
\node[] (A2) at (0.8,-1.0) {$-\mathbf{n_2}$};
\node[] (A2) at (6.5, 4.7) {$z_2$};
\node[] (A2) at (6.5, -1.5) {$z_1$};
\node[] (A) at (1.7,-1.8) {$-\mathbf{n}$};
\draw[->, thick]  (B) -- (1.8, -0.55);
\draw[->, thick]  (B) -- (2.5, -1.875);
\node[draw,circle] (gal) at (6, 4.0) {{\textsc{Ly}$\alpha$}};
\node[] (zm) at (6, 1.06) {};
\fill (zm) circle [radius=2pt];
\node[] (test) at (0.0, 5.0){};
\draw[dashed, -, thick] (B) -- (zm) node [midway, above, sloped] (TextNode) {$\chi_{m}$};;
\node[draw, circle] (h1) at (6, -1.0) {{\textsc{Q}}};
\draw[dashed, -, thick]  (B) -- (gal);
\draw[dashed, -, thick] (B) -- (gal) node [midway, above, sloped] (TextNode) {$D_\alpha$};;
\draw[dashed, -, thick]  (B) -- (h1);
\draw[dashed, -, thick] (B) -- (h1) node [midway, below, sloped] (TextNode) {$D_\mathrm{Q}$};;
\draw[dashed, -, thick]  (h1) -- (gal);
\draw[->, thick]  (B) -- (2.2, -1.19466666667);
\node[] (mm) at  (8.47, 2.52553333333) {};
\node[] (AA) at  (9.52, 1.04853333333) {};
\node[] (m) at (7.47, 1.9232) {};
\node[] (A2) at (8.52, 0.4552) {};
\node[] (rp) at (8.5, 0.483333333333) {};
\node[] (rt) at (7.47, 1.427066666667) {};
\fill (rp) circle [radius=2pt];
\node[] (rp1) at (6.0, -2.0) {};
\node[] (rp2) at (8.5, -0.5166666666670001) {};
\draw[dashed, -, thick]  (B) -- (m);
\draw[dashed, -, thick]  (gal) -- (rp);
\draw[dashed, -, thick]  (h1) -- (rp);
\draw[->, thick]  (h1) -- (6, 1.0);
\node[] (N) at (6.3, 0.0) {$\mathbf{N}$};
\draw [decorate,decoration={brace,amplitude=10pt,mirror,raise=4pt},yshift=0pt]
(gal) -- (h1) node [black,midway,xshift=-0.8cm] {\footnotesize $r$};
\draw [decorate,decoration={brace,amplitude=10pt,mirror,raise=4pt},yshift=0pt]
(8.5, 0.43) -- (gal) node [black,midway,xshift=0.8cm, yshift = 0.4cm] {\footnotesize $r_\perp$};
\draw [decorate,decoration={brace,amplitude=10pt,mirror,raise=4pt},yshift=0pt]
(h1)-- (8.5, 0.5) node [black,midway,xshift=0.4cm, yshift = -0.7cm] {\footnotesize $r_\parallel$};
\draw[thick] ([shift=(30:1cm)]6,1.06) arc (30:90:1cm);
\node[] (b) at (6.6,2.3) {$\beta$};
\draw[thick] ([shift=(14:2.9cm)]0,-2.5) arc (14:31:2.9cm);
\draw[thick] ([shift=(14:3cm)]0,-2.5) arc (14:31:3cm);

\draw[thick] ([shift=(14:5cm)]0,-2.5) arc (14:47:5cm);

\node[] (X) at (3.2,-1.25) {$\theta/2$};
\node[] (X) at (4.4,0.5) {$\theta$};

\end{tikzpicture}
\end{center} 
 \caption {Illustration of the coordinate system adopted in our work, in agreement with Ref.~\cite{Bourboux:2017cbm}. 
}
\label{coord}
\end{figure}

The coordinate system described above can be formulated in terms
of the separation between the two tracers $r = \left| D_\alpha \bn_2 - D_\text{Q} \bn_1 \right|$ and the coordinate $\cos{\beta} \equiv -\bn \cdot \mathbf{N}$, where $\mathbf{N}$ and $-\bn$ are the unit vectors pointing in the direction of the distance between the two tracers and in the direction of the mean redshift, respectively. 
In this notation, the separation between the two tracers is
positive-definite, while the reference angle $\beta$ rotates from $0$ to $2 \pi$.
In terms of $r$ and $\beta$, the traverse and
parallel separations read 
\begin{equation}
r_\perp = r \sin{\beta}, \qquad r_\parallel = r \cos{\beta}.
\end{equation}
In Fig.~\ref{coord} we draw a scheme of the system under investigation and we show the reference coordinates. 

The measured correlation function can be expressed as a sum of several contributions:
\begin{equation}
\xi_\text{tot} = \xi_{\text{Q}, \alpha} + \xi_\text{Q, met} + \xi_\text{Q, HCD} + \xi_\text{TP}, \label{mod1}
\end{equation} 
where  $\xi_{\text{Q}, \alpha}$ is the cross-correlation between quasars and Lyman-$\alpha$ absorption, $\xi_\text{Q, met}$ and $\xi_\text{Q, HCD}$ are the contaminants due to metals and unidentified High Column Density absorbers (HCDa), respectively, and $\xi_\text{TP}$ models the Transverse Proximity effect (TP).
The cross-correlation for Q-$\alpha$ and the other absorbers are estimated as the Fourier transform of the power spectrum, i.e.~by assuming the flat-sky approximation:
\be
\xi \left( r, \mu \right)  = \int \frac{d k k^2 d \mu_k}{\left( 2 \pi \right)^2} P \left( k, \mu_k \right) J_0\left(  k r \sqrt{1 - \mu^2} \sqrt{1 - \mu_k^2}\right) e^{-i r k \mu \mu_k},
\ee
where $J_0$ is the 0-order Bessel function, $\mu = \cos{\beta}$, while $k$ and $\mu_k$ are the Fourier space coordinates $k = \left| \mathbf{k} \right|$ and
$\mu_k= - \bn \cdot \hat\bk$. 
We omitted the dependence on the mean redshift for simplicity. 

The Q-$\alpha$ cross-spectrum includes linear density perturbations, linear redshift-space distortion (RSD) and non-linear corrections\footnote{In Ref.~\cite{Bourboux:2017cbm} a non-linear damping of the Baryon Acoustic Oscillations feature is included in the model. In our work, we neglect this effect.}:
\begin{equation}
P^{\text{Q}, \alpha}(k, \mu_k) = P_\text{L} (k, \mu_k) \sqrt{V_\text{NL}(k, \mu_k)} \sqrt{F_\text{NL} (k, \mu_k)} G(k, \mu_k).
\label{eq:Pqa}
\end{equation}
The linear cross-spectrum $P_\text{L} (k, \mu_k)$ is
\begin{equation}
P_\text{L} (k, \mu_k) = P_\text{m} (k) 
b_\text{Q}b_\alpha (1+f/b_\mathrm{Q}\mu_k^2) (1 + \beta_\alpha \mu_k^2)
\label{linPk}
\end{equation}

where $P_\text{m} (k)$ is the linear matter power spectrum, 
$b_\text{Q}$ and $b_{\alpha}$ are the clustering biases of the two tracers, while $f$ and $\beta_\alpha$ are the growth factor and the RSD parameter for the Lyman-$\alpha$, respectively. 
While the QSOs bias is assumed to be a function of the redshift alone, Ly$\alpha$ bias is scale dependent due to fluctuations in the ionizing UV radiation. The scale dependence 
is modelled as follows \cite{Gontcho:2014nsa, Bourboux:2017cbm}:
\be
b_\alpha(k) = \bar{b}_\alpha + b_{\Gamma}\frac{W(k\,\lambda_\text{UV})}{1-2/3 W(k\,\lambda_\text{UV})},
\ee
where $b_{\Gamma}$ is a parameter of the model and $\lambda_\text{UV} = 300 h^{-1}$Mpc is the mean free path of the UV photons. 

Note that the parameter $\beta_\alpha$ for Ly$\alpha$ is scale-dependent as well. Indeed,
it can be expressed in terms of the growth factor as $\beta_\alpha = b_\text{v}f/b_\alpha$, where  
$b_\text{v}$ denotes the Ly$\alpha$ velocity bias (see e.g. Refs.~\cite{uros12,arino15}). 
Therefore, $\beta_\alpha$ is scale-dependent through the clustering bias $b_\alpha$.
However, the RSD contribution 
to the power spectrum is scale-independent, since the growth factor depends only on redshift in General Relativity.

Furthermore, in Eq.~\eqref{eq:Pqa}, there are three additional terms describing: non-linear corrections due to quasar velocities ($V_{\rm NL}$), non-linear effects of the Lyman-$\alpha$ forest ($F_{\rm NL}$)\footnote{This includes temperature and thermal pressure smoothing, as well as non-linear correction to peculiar velocities.}, and a correction due to finite size of the observed ($r_\parallel,r_\perp$) bins ($G_{\rm NL}$). All of these three functional forms are aimed at small-scale corrections, and will have little impact on our final result. We model them as in Ref.~\cite{Bourboux:2017cbm}.

Metals that have a transition close to the Ly$\alpha$ transition are present in the integalactic medium and contribute to the observed correlation function.
This contamination can be modelled similarly to  Ly$\alpha$-QSOs cross-correlation, with the complications that the observed coordinates $\Delta \theta$ and
$z$ are converted into the real space coordinates $r_\parallel$ and $r_\perp$ assuming that the absorption is due to Lyman-$\alpha$ transition. In order to take into account the different
source rest-frame wavelength of the metal absorption, the coordinate $r_\parallel$ needs to be appropriately rescaled. 
Since the metals' contribution peaks at the physical separation between the two tracers close to zero and the peak is displaced in one direction from $r_\parallel = 0$, it naturally introduces an asymmetric feature in the correlation function. 

The metal transition included in the BOSS model and
the corresponding correction to $r_\parallel$ are summarized in Ref. \cite{Bourboux:2017cbm}, table 3. As in Ref. \cite{Bourboux:2017cbm}, we fix $\beta_{\rm m} = 0.5$ for all metal species.
 
The remaining two effects modelled in Eq.~\eqref{mod1} are corrections due to unidentified high column density (HCD) systems correlating with quasar positions, and the effect of quasar radiation affecting ionization close to itself -- characterized by transverse proximity effect. To first order approximation we model both of them as in~\cite{Font-Ribera:2013wce, Bourboux:2017cbm}.

The Ly$\alpha$ parameters are derived from the best-fit values
of the auto-correlation data alone, while the parameters
that do not affect the auto-correlation of the Ly$\alpha$ are fixed to the fitted values from the cross-correlation data (see
table 4 in Ref. \cite{Bourboux:2017cbm}). 

\begin{figure}
\centering
    \begin{subfigure}[b]{0.485\textwidth}
        \includegraphics[width=\textwidth]{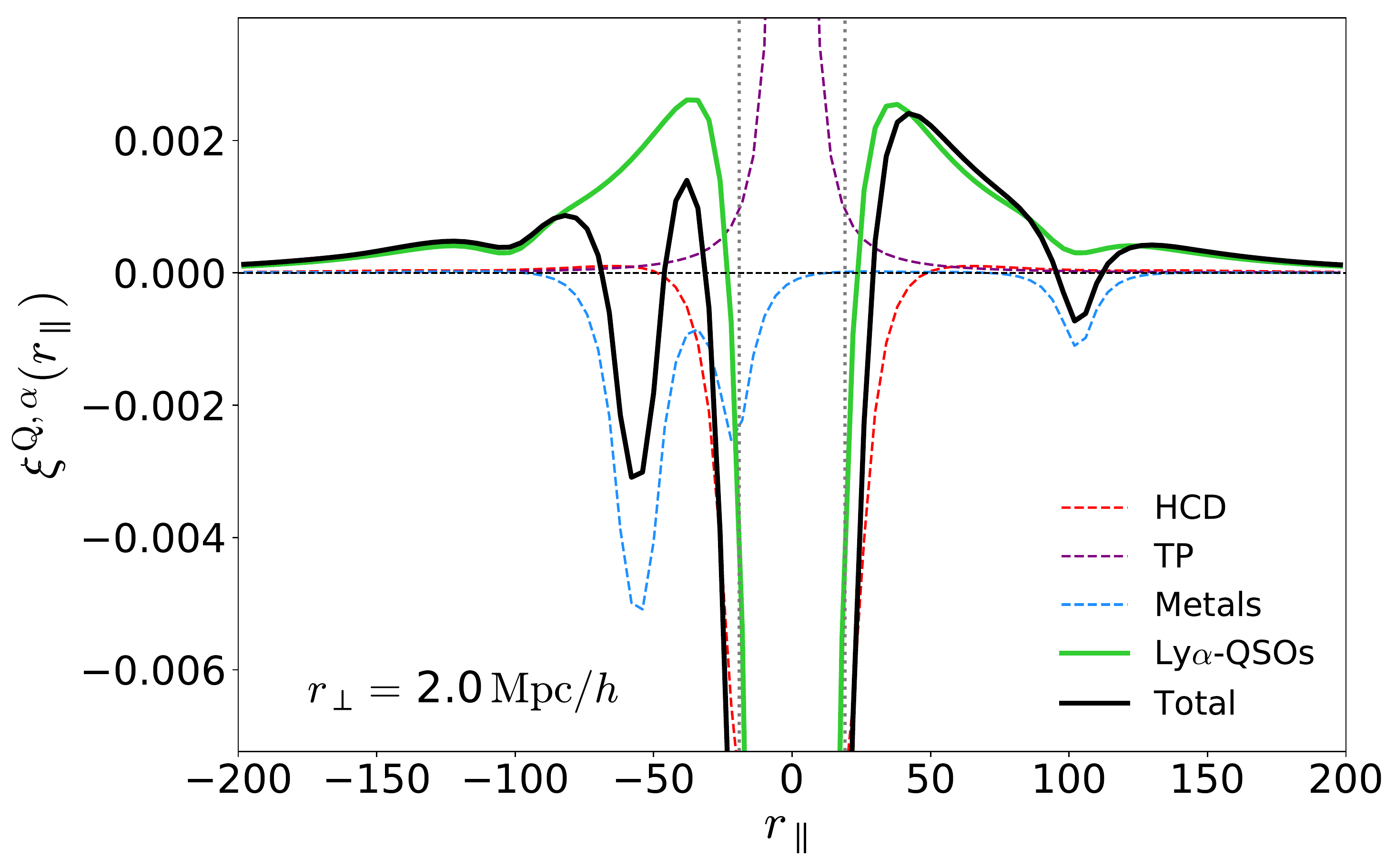}
        \caption{$r_\perp = 2 \text{Mpc}/h$}
        \label{fig:rt2l}
    \end{subfigure}
    \begin{subfigure}[b]{0.5\textwidth}
        \includegraphics[width=\textwidth]{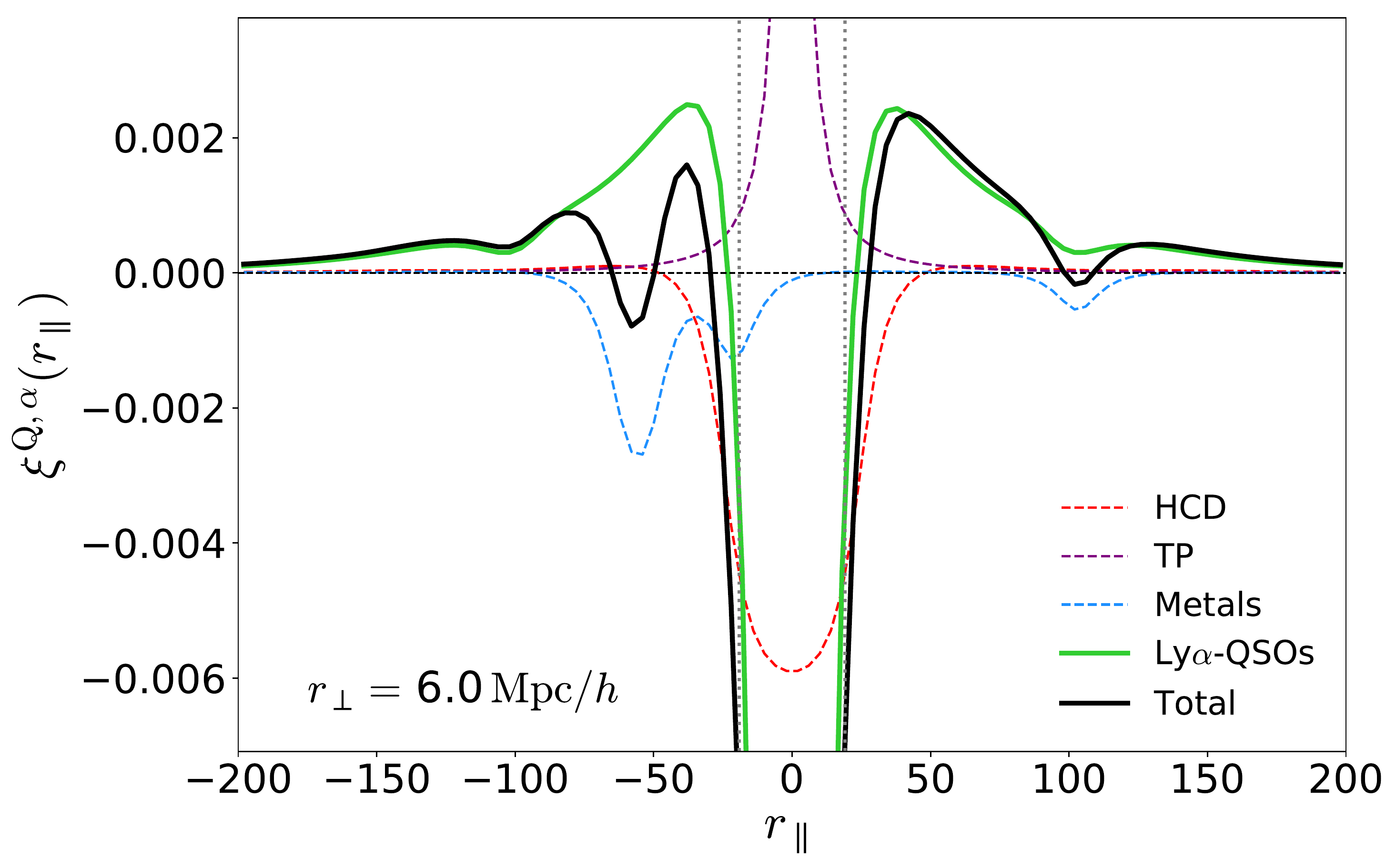}
        \caption{$r_\perp = 6 \text{Mpc}/h$}
        \label{fig:rt6}
    \end{subfigure}
    \caption{Model for the observed cross-correlation, as a function of the longitudinal distance between the tracers. Different colors outline the different contributions,
    the black line represents the sum of all of them. Vertical dotted lines represent a typical cut scale in $r_\parallel$, estimated assuming $r_\mathrm{cut} = \sqrt{r^2_{\parallel, \text{cut}} + r_\perp^2} = 20\,\mathrm{Mpc}/h$. For $r < r_\mathrm{cut}$ contamination of small scales effects not included in the model might be not negligible. Therefore, scales below this threshold are excluded in typical data analysis.
    Fig. \ref{fig:rt2l} refers to $r_\perp = 2\, \text{Mpc}/h$, Fig. \ref{fig:rt6} refers to $r_\perp = 6 \,\text{Mpc}/h$.}
    \label{fig:BOSS}
\end{figure}

In Fig.~\ref{fig:BOSS} we show the different contributions to the observed cross-correlation. The HCD contamination overcomes the Ly$\alpha$-QSOs cross-correlation on scales
$r_\parallel \lsim 60 \text{Mpc}/h$. Metal contamination introduces an asymmetry in the cross-correlation. 
\begin{figure}
\centering
    \begin{subfigure}[b]{0.485\textwidth}
        \includegraphics[width=\textwidth]{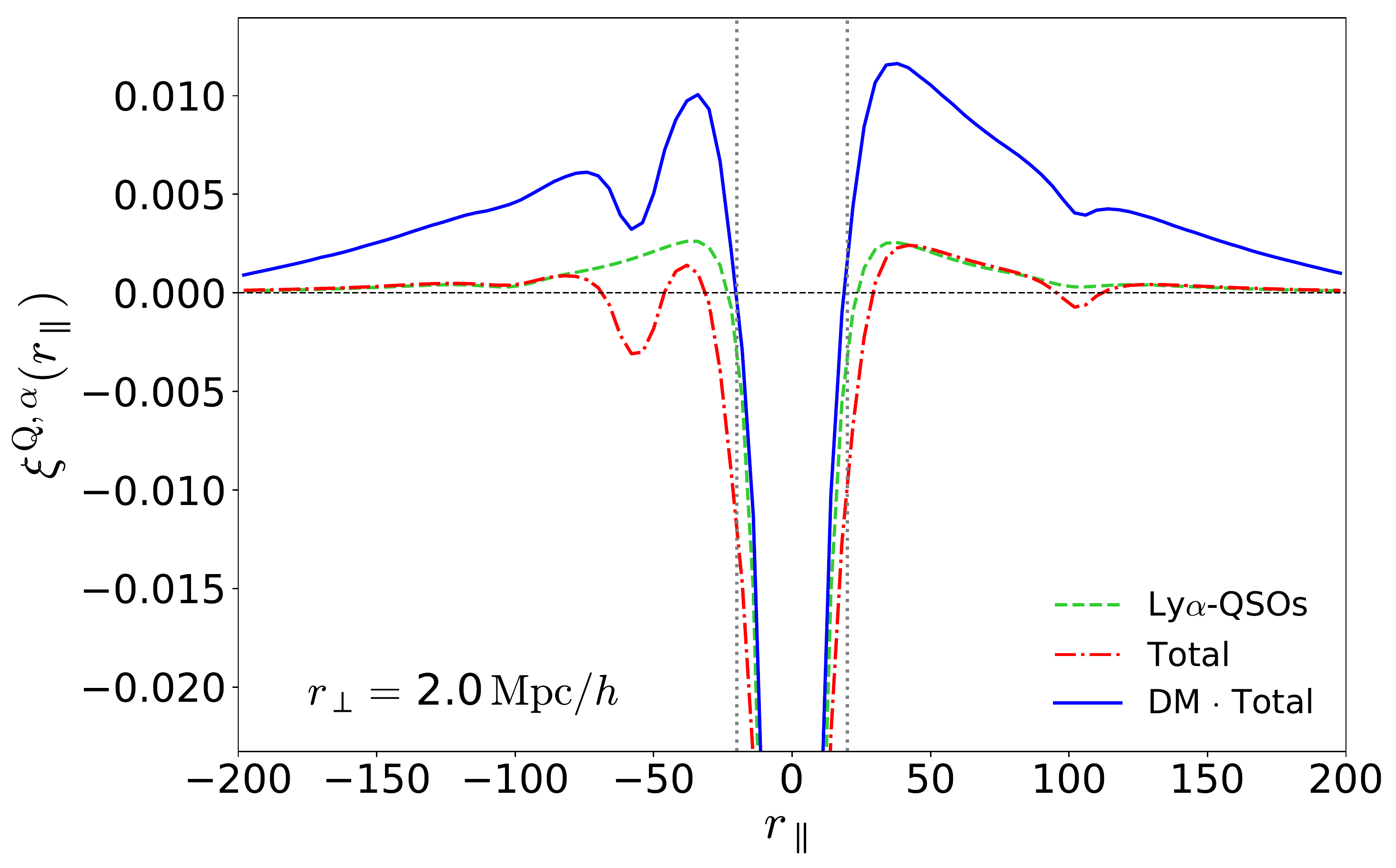}
        \caption{$r_\perp = 2 \text{Mpc}/h$}
        \label{fig:rt2l}
    \end{subfigure}
    \begin{subfigure}[b]{0.5\textwidth}
        \includegraphics[width=\textwidth]{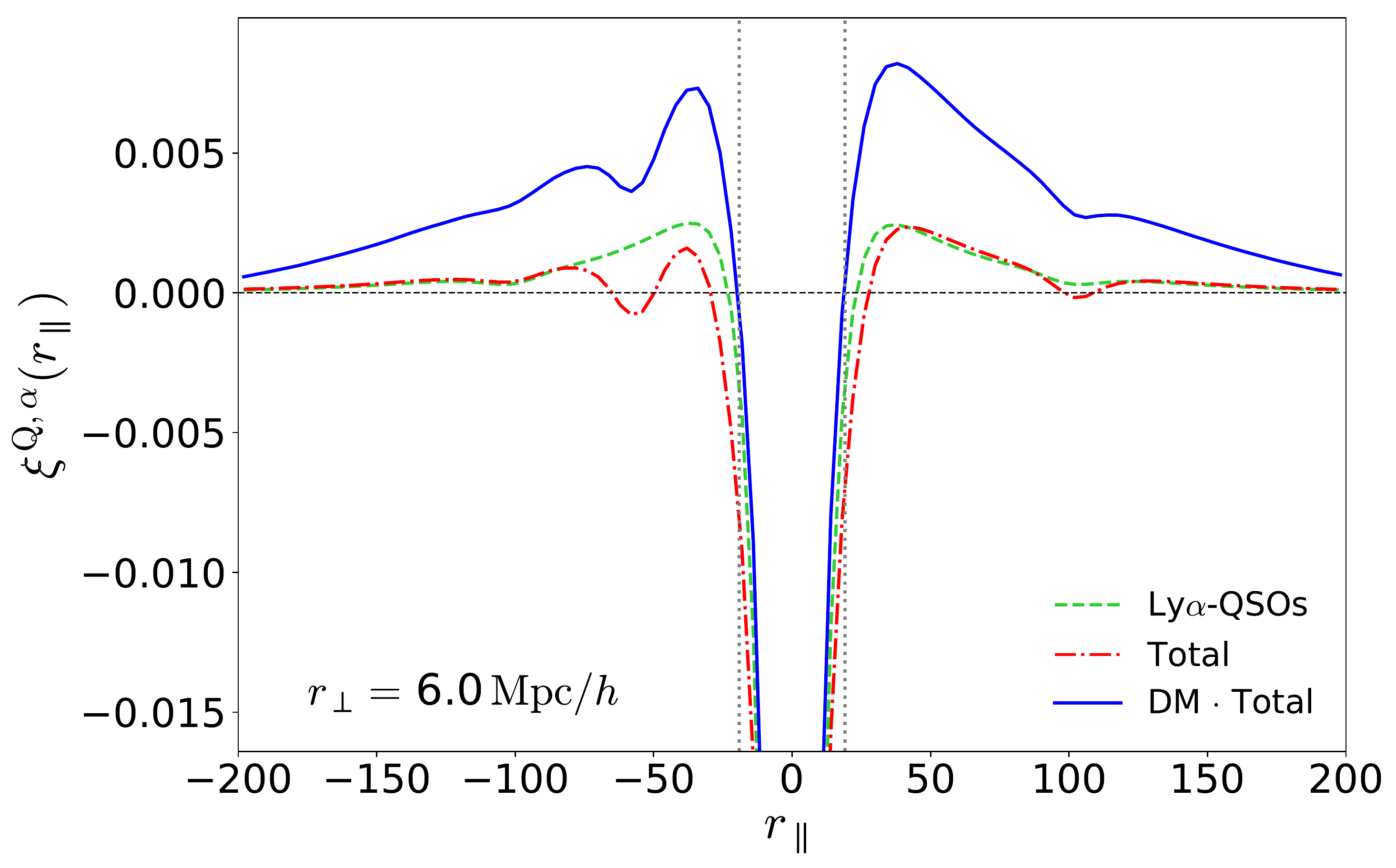}
        \caption{$r_\perp = 6 \text{Mpc}/h$}
        \label{fig:rt6}
    \end{subfigure}
    \caption{QSOs-Ly$\alpha$ cross-correlation (green dashed lines), total cross-correlation with (blue continuous lines) and without (red dashed-dotted lines) distortion matrix. 
    Vertical dotted lines represent a typical cut scale $r_\mathrm{cut} = \sqrt{r^2_{\parallel, \text{cut}}  + r_\perp^2} = 20\,\mathrm{Mpc}/h$.
    Left panel refers to $r_\perp = 2\, \text{Mpc}/h$, right panel
    refers to  $r_\perp = 6\, \text{Mpc}/h$.}
    \label{fig:BOSS2}
\end{figure}
On top of the metals, HCD, and proximity effect corrections, there is a mixing effect in the correlation of quasars-pixels which is modelled through the distortion matrix. 
The distortion matrix summarizes the imperfect knowledge of the quasar intrinsic continuum, and mean transmission variations. Because the Lyman-$\alpha$ absorption features are removing a lot of the intrinsic spectrum blue-wards of the Lyman-$\alpha$ transmission line, it makes it hard to estimate the exact shape of the quasar continuum. While there are strengths and weaknesses of different methods used, all of them are deficient in characterizing the (long-wavelength) fluctuations of the continuum, that might be misinterpreted as large-scale Lyman-$\alpha$ flux fluctuations. To correct for this effect in a specific survey, the BOSS collaboration \cite{Bautista:2017zgn} removes the bias introduced due to their choice of fitting method. This results in exact vanishing first momentum of the flux fluctuation field, largely removing the bias that the continuum fitting (and mean transmission) introduces.

Since the distortion matrix affects the correlation function mainly at large scales, it is a primary instrumental effect that could mask or distort the signal from relativistic corrections considered in this work. However, the distortion matrix is an effect that can be calibrated through the specific survey (see for instance~\cite{Bautista:2017zgn}).
Therefore, the measured correlation function will be a linearly distorted version of the one modelled in~\eqref{mod1}
\be
\xi_\text{obs}^i = {\text{D}}_{ij} \xi_\text{tot}^j 
\ee
where $\text{D}$ is the distortion matrix (see Ref.~\cite{Bourboux:2017cbm} for details) and the indices $i,j$ run over the different bins on the plane $(r_\perp, r_\parallel)$. 

In Fig.~\ref{fig:BOSS2} we show the impact of the distortion matrix on the total cross-correlation function, for the bins corresponding to the configuration $r_\perp = 2 \text{Mpc}/h$ and $r_\perp = 6  \text{Mpc}/h$.
We see that the distortion matrix significantly boosts the signal for small transverse pair separations.

\begin{figure}
\centering
    \begin{subfigure}[b]{0.49\textwidth}
        \includegraphics[width=\textwidth]{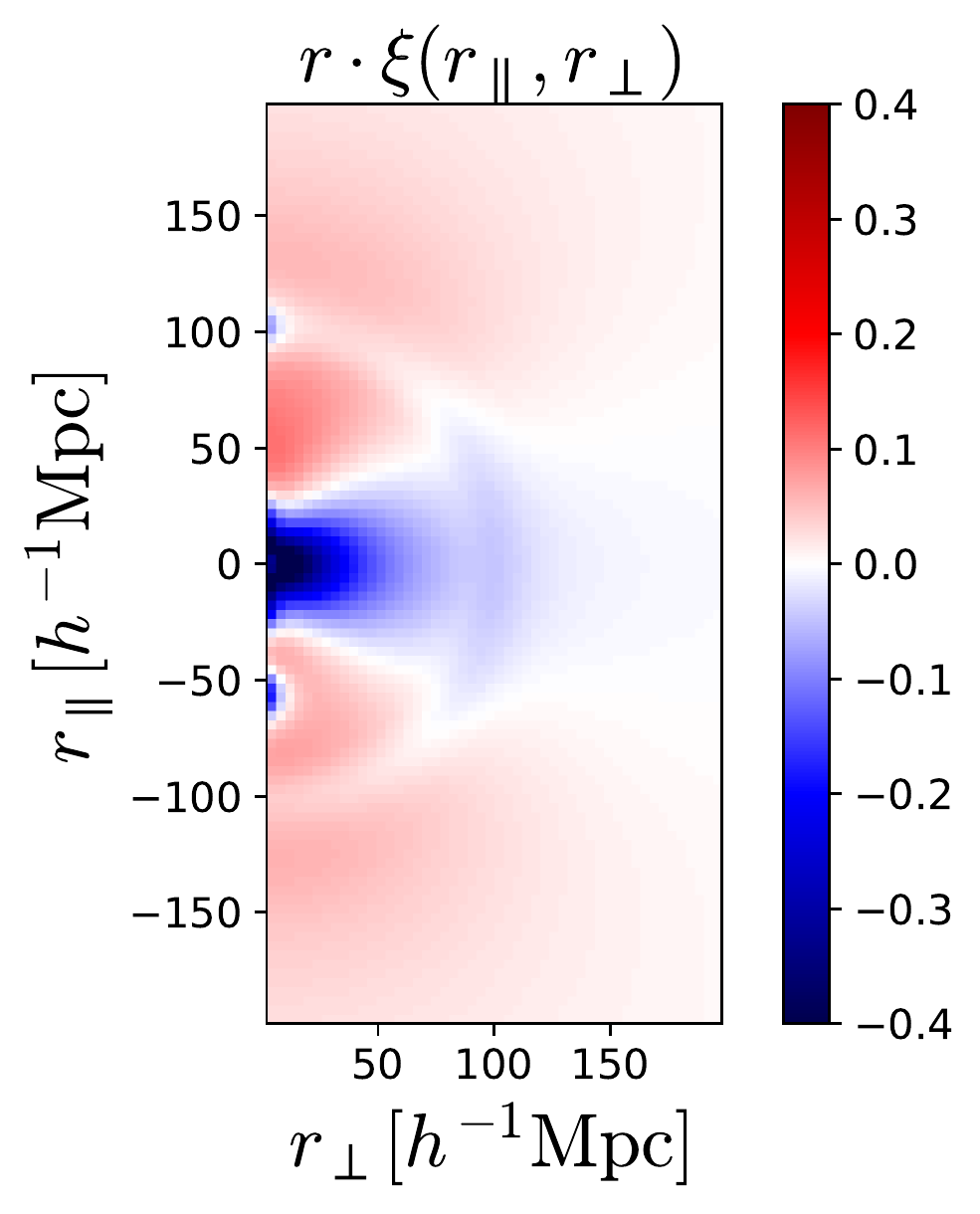}
        \caption{}
    \end{subfigure}
    \begin{subfigure}[b]{0.49\textwidth}
        \includegraphics[width=\textwidth]{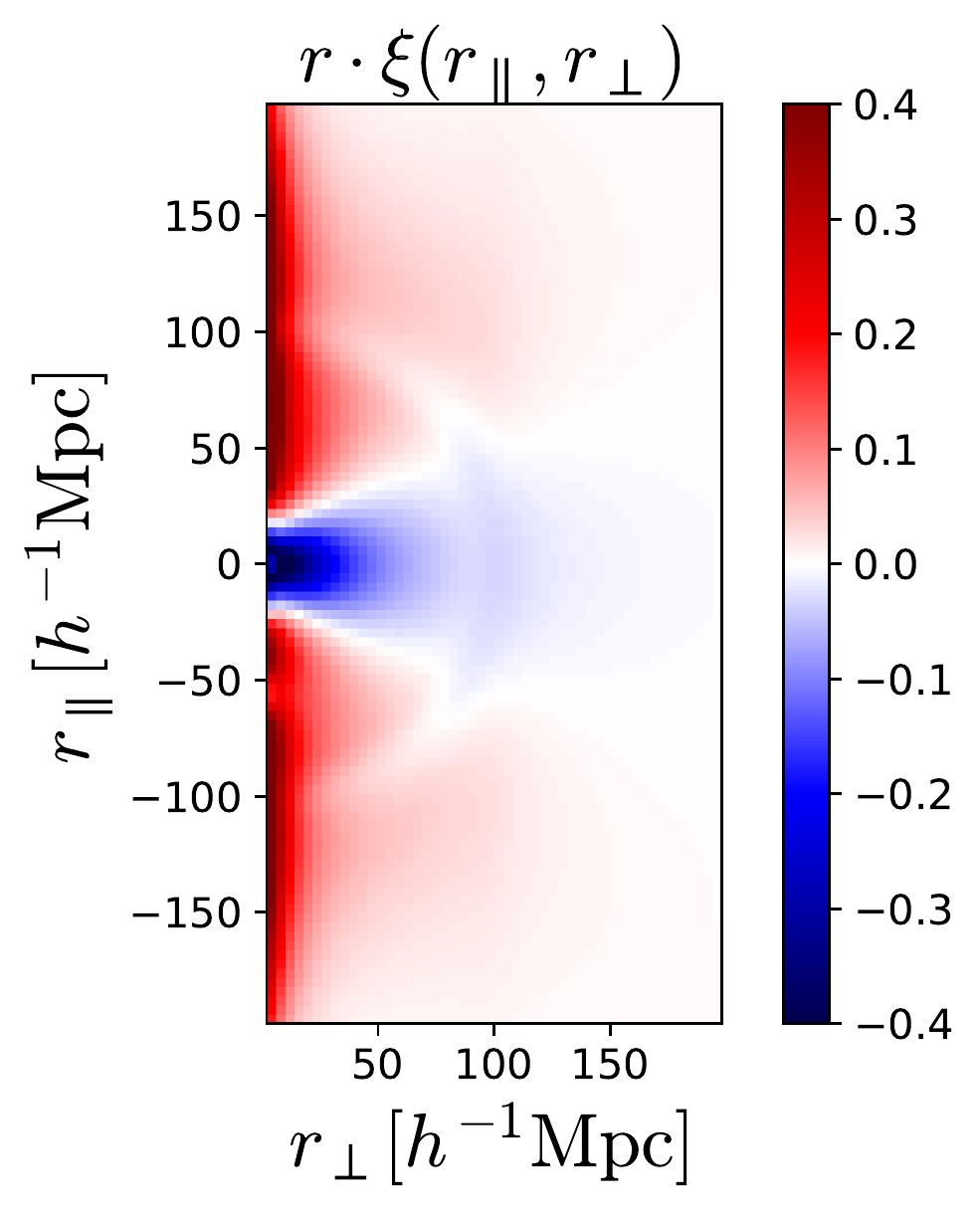} 
        \caption{}
    \end{subfigure}
    \caption{QSOs-Ly$\alpha$ cross-correlation in the 
    $(r_\perp, r_\parallel)$ plane as modelled in Ref.~\cite{Bourboux:2017cbm} without (left panel) and with
    (right panel) the distortion matrix. The correlation function has been sampled in bins with size $\Delta r_\parallel = \Delta r_\perp = 4\, \text{Mpc}/h$.}
    \label{fig:BOSS3}
\end{figure}

\section{Relativistic effects in the Ly$\alpha$-quasar cross-correlation function}
\label{sec:rel-Eff}
The model outlined in the previous section neglects all the relativistic effects, beyond the standard redshift space distortion, that contribute to the cross-correlation between Lyman-$\alpha$ absorption and quasar $\xi_{Q,\alpha}$.
In a relativistic framework we need to properly account for the fact that we observe only galaxies, or in general biased tracers of the underlying dark matter field, which lie on our past light cone. In this view, the 3-dimensional mapping between real and RSD coordinates represents only a first approximation on small scales (with respect to the Hubble scale $\HH^{-1}$). The light-cone is defined through the photon propagation and, in a clumpy universe, this is affected as well by lensing, integrated Sachs-Wolfe and time-delay effects. Moreover the source motions induce further corrections, such as Doppler effects. The theoretical description of LSS in a relativistic framework has been investigated extensively in the last decade. The galaxies number counts including all the relativistic effects has been derived within linear theory in Refs.~\cite{yooA,yooB,challinor_lewis,bonvin_durrer,DiDio:2016ykq,Durrer:2016jzq} and then extended to other observables (see Ref.~\cite{Hall:2012} for HI mapping and Ref.~\cite{Irsic:2015} for Ly$\alpha$ flux absorption) and to higher orders in perturbation theory (see Refs.~\cite{Yoo:2014sfa,Bertacca:2014dra,DiDio:2014lka} for second and Refs.~\cite{Nielsen:2016ldx,Jalivand:2018vfz,DiDio:2018zmk} for third orders).
At linear order, the gauge-invariant relativistic number counts read as follows (by following the notation of Refs.~\cite{bonvin_durrer,CLASSgal})
\bea \label{counts1}
\Delta(\bn, z) &=& b \delta  + \HH^{-1} \partial_\chi \ndv 
 + (5s - 2) \int_0^{\chi} \frac{\chi - \chi'}{2 \chi \chi'} \Delta_{\Omega} (\Phi + \Psi) d\chi' \nonumber  \\
&&+ \left(\frac{\dot \HH}{\HH^2} + \frac{2-5 s}{\chi\HH} + 5 s - f_{\text{evo}}
\right)\ndv  \nonumber \\
&&+\left( f_{\text{evo}}  - 3\right)\HH V +(5s - 2) \Phi + \Psi + \HH^{-1} \dot \Phi +\frac{2-5s}{\chi} \int^{\chi}_0 d\chi' (\Phi + \Psi) \nonumber \\
&& +\left(\frac{\dot \HH}{\HH^2} + \frac{2-5 s }{\chi\mathcal{H}}+5s  - f_{\text{evo}}\right)\left(\Psi + \int^{\chi}_0 d\chi' \left(\dot\Phi + \dot\Psi \right)\right) \, ,
\eea
where $\delta$ denotes the dark matter density fluctuation in comoving gauge, $\ndv = \bn \cdot \bv$ is the velocity along the line-of-sight and $\bv$ is the peculiar velocity in Newtonian gauge. The metric perturbations $\Phi$ and $\Psi$ correspond to the two Bardeen potentials and $V$ is the velocity potential defined through $\bv = - {\bf\nabla}V $. We denote the partial derivative with respect to the conformal time $t$ with a dot and the integrals run along the past light-cone. The comoving Hubble parameter is defined through $\HH= \dot a/a$ and $\chi$ is the comoving distance to redshift $z$. 
In order to relate dark matter perturbations to galaxies (or other discrete biased tracers of the matter field) we have introduced three different bias parameters: a linear galaxy bias $b$, a magnification bias
\be
s= - \frac{2}{5} \left.\frac{\partial \ln \ \bar n \left( z , \ln L \right)}{\partial \ln L}\right|_{\bar L}
\ee
where $\bar L$ denotes the threshold luminosity of the survey and $\bar n$ the background number density, and an evolution bias
\be
f_{\rm evo}= 3 - \left( 1 +z \right) \frac{d \ln \ \bar n}{dz}
\ee
which describes the deviation from number conservation of sources in a comoving volume.

In our work we are interested in the cross-correlation between QSOs and Ly$\alpha$, therefore we need to describe also the Ly$\alpha$ flux absorption in a relativistic framework. By following Ref.~\cite{Irsic:2015} this can be written as
\bea 
\delta_F \left( \bn , z \right) &=&  b_\alpha \delta^\text{sync} + b_v
\HH^{-1} \partial_\chi \ndv  + b_R \!
\left[ - \!\left( \!2 + \frac{\dot \HH}{\HH^2} - f_\text{evo}^{\alpha} \!\right)\frac{\delta z}{1+z} \right.
\nonumber \\
&& 
\label{Lya_Delta}
\hspace{2cm}
\left.
+\ndv + \Psi   + \HH^{-1}\dot \Phi  + \left( f_\text{evo}^{\alpha} -  3 \right)  \HH V \!\right] \, ,
\eea
where 
\be
 \frac{\delta z}{1+z}  = - \left(   \Psi + \ndv +  \int_0^{\chi} \left( \dot \Psi + \dot \Phi \right) d\chi'  \right) 
\ee
is the linear redshift perturbation. The velocity bias $b_v$ of the Lyman-$\alpha$ forest arises because of the non-linear transformation between the observable (flux) and the physical quantity tracing the Lyman-$\alpha$ forest structure (optical depth). Indeed, in optical depth, the velocity bias would be equal to 1 (disregarding small corrections due to velocity gradients across the absorbers \cite{Arinyo-i-Prats:2015vqa, Irsic:2018hhg}), and thus resembles galaxy surveys.
We thus assume that the bias of the relativistic corrections $b_R$ traces the same mapping as redshift space distortions term, and should be equal to velocity bias $b_R= b_v$. These assumptions are valid at the large scales we are investigating here.
If we compare the expression for discrete number counts, eq.~\eqref{counts1}, with the equivalent one for Ly$\alpha$ flux absorption, eq.~\eqref{Lya_Delta}, we remark that the latter is not affected by lensing magnification. Indeed, by being a smooth field, the change in the solid angle $d\Omega$ and the change in the observed flux compensate each other because of the surface brightness conservation. Lensing effects appear only at second order in terms of deflection angle~\cite{Yoo:2014sfa,Bertacca:2014dra,DiDio:2014lka,DiDio:2016kyh}. It is worth noticing that, while the Newtonian contributions enter in the same form for QSO and Ly$\alpha$, the relativistic corrections have a different form, even by suppressing the magnification contribution in the QSOs expression, i.e.~setting $s=0.4$.
Indeed in this case, the Doppler effect appears in the two observable as follows
\bea
\left(\frac{\dot \HH}{\HH^2} + 2 - f_{\text{evo}}
\right)\ndv  \subset \Delta \left( \bn, z \right) \, ,
\\
b_R \left(\frac{\dot \HH}{\HH^2} +3 - f^{{\rm Ly}\alpha}_{\text{evo}}
\right)\ndv  \subset \delta_F \left( \bn, z \right) \, .
\eea
The first expression agrees with the Doppler contribution to HI mapping observable~\cite{Hall:2012}. The different numerical factors between the two sources is a consequence of their background evolutions. While the background temperature scales as $ \bar T_b = n_{\rm HI}/ (\mathcal{H} (1+z)^2 )$, in case of Ly$\alpha$ the optical depth evolves as $\bar \tau =n_{\rm HI}/ (\mathcal{H} (1+z) )$. \footnote{
The background brightness temperature and optical depth have a different redshift evolution $\bar \tau \propto \left( 1+ \bar z \right) \bar T_b$. By replacing the background redshift $\bar z$ with the observed redshift $\bar z= z - \delta z$, we have
\be
\bar \tau \propto \left( 1 + z - \delta z \right) \bar T_b =
\left( 1- \frac{\delta z}{1+z } \right) \left( 1+ z \right) \bar T_b
=  \left( 1 + v_\parallel  \right) \left( 1+ z \right) \bar T_b \, .
\ee
Therefore in the Lyman-alpha observable we have an additional contribution proportional to the longitudinal velocity $\ndv$.
}
 
To estimate the amplitude of the different contributions to the 2-point function we work in the weak field approximation, namely we write the previous expressions in terms of the expansion parameter $\HH/k$, and we consider the relation between the matter and metric perturbations as predicted by the Poisson equation
\be
\delta \sim \left( k/\HH \right)^2 \Phi
\ee
and with the velocity through the Euler equation
\be
v \sim \left( k/\HH \right) \Phi \, .
\ee
Within this approximation, the first line of eq.~\eqref{counts1} appear to 0-order, the second line to first order and the remaining lines to second order in the weak field parameter $\HH/k$. Analogously, in eq.~\eqref{Lya_Delta} the first two terms are at the leading order, while the terms in square brackets involving the radial velocity $\ndv$ are to first order and all the others to second order in the expansion parameter $\HH/k$. Therefore the first term beyond the standard Newtonian approximation (the terms of the 0-order in our weak field expansion) are suppressed by a factor $\HH/k$. This is of the order $\sim 6 \times 10^{-3}$ at the BAO scale and therefore we can safely neglect any further corrections starting from the order $\left( \HH/k \right)^2$. The leading relativistic corrections for cross-correlation of different probes or multitracer analysis, are at the order $\HH/k$ and proportional to the velocity along the line of sight. These terms are commonly known as Doppler corrections, and they have been investigated for different probes, see e.g. Refs.~\cite{mcdonald:2009,Yoo:2012se,Bonvin:2013,alonso15,Fonseca:2015laa,Irsic:2015,Bonvin:2015,Bonvin:2016, Hall:2016, Lepori:2017twd,Breton:2018wzk,Bonvin:2018ckp}.

In our work we aim to quantify the impact of the Doppler correction in the measured 3-dimensional correlation function between Ly$\alpha$ and QSOs.

\section{Doppler, z-evolution and wide-angle signal}

\label{sec:rel-Eff-2}

The relativistic cross-correlation function between
QSOs and Ly$\alpha$ is 
\begin{equation}
\xi^{\text{Q}, \alpha}_\text{rel}(z_1, z_2, \theta) =  \mean{\Delta_\text{Q}(\mathbf{n}_1, z_1)\Delta_\alpha(\mathbf{n}_2, z_2)}, \qquad \cos{\theta} \equiv \mathbf{n_1}
\cdot \mathbf{n_2}. \label{full_cross}
\end{equation}
In our work, we neglect the gravitational lensing contribution\footnote{The impact of lensing contamination for the Ly$\alpha$-QSOs cross-correlation is discussed in Appendix \ref{ap-lens}.}
and all the terms of the order $\left( \HH/k \right)^2$ in
the quasars number counts and Ly$\alpha$ flux fluctuations.
Considering that these effects are negligible, 
we can schematically write the relativistic correlation function
as a sum of the following terms:
\be
\xi_\text{rel} = \xi_\text{stand-flat} + \xi_\text{wa} 
+ \xi_\text{evo} + \xi_\text{Doppler}.
\ee
The 'stand-flat' contribution includes the cross-correlation
of the density and RSD for the two tracers within flat-sky approximation, with
all redshift-dependent factors evaluated at the effective
redshift of the survey. We discussed this contribution in
detail in Sec.~\ref{boss-mod}. 
The 'wa' term includes the corrections from density and
RSD to the flat-sky approximation, while 'evo' 
models the redshift evolution of the biases and the growth factor.
In our notation, the 'Doppler' term includes the cross-correlation 
of density and RSD with the Doppler correction for the two tracers.
The cross-correlation of the Doppler corrections for QSOs and 
Ly$\alpha$ is neglected, since it contributes at the order
$\left( \HH/k \right)^2$.

The wide-angle and evolution corrections are currently neglected
in the Ly$\alpha$-QSOs cross-correlation analysis. For 
pair separation $r \ll \chi_\text{m}$, being $\chi_\text{m}$ the comoving distance to the mean redshift of the survey, they can be safely neglected. In the current state-of-the-art analysis \cite{Bourboux:2017cbm}, 
for the largest separation of the two tracers we have
$r/\chi_\text{m} \approx 7 \times 10^{-2}$. Therefore, we include
in the wide-angle and evolution terms the lowest-order
correction in $r/\chi_\text{m}$.
These contributions are given by \cite{Bonvin:2013}
\begin{align}
\xi_\text{wa}(r, \mu) &= \left(b_\text{Q}b_\text{v} - \bar{b}_\alpha\right) \left[P_1(\mu) - P_3(\mu)\right]  \frac{2}{5}f \frac{r}{\chi_\text{m}} \mu_2(r)  \\
&- \left[P_1(\mu) -  P_3(\mu)\right]\frac{2}{5} \frac{r}{\chi_\text{m}}f \mu_{2, \alpha}(r)\notag \\
\xi_\text{evo}(r, \mu) &= \Biggl[ - (b_\text{Q} (b_\text{v} f)'-
\bar{b}_\alpha f')\left(\mu_0(r) - \frac{4}{5} \mu_2(r)\right)+ \\
&+ f(b'_\text{Q} b_\text{v} - b'_\alpha)\left(\mu_0(r) - \frac{4}{5} \mu_2(r)\right) - 3\left(b_\text{Q}b'_\alpha -  b'_\text{Q}b_\alpha\right) \mu_0(r) \Biggr]\frac{r}{6} P_1(\mu) \notag \\
&+\Bigl[b_\text{Q}(b_\text{v} f)' - \bar{b}_\alpha f' - (b'_\text{Q}b_\text{v} - b'_\alpha) f \Bigr] \frac{r}{5} \mu_2(r) P_3(\mu) \notag \\
&+f'\left[\mu_{0, \alpha}(r) - \frac{4}{5} \mu_{2, \alpha}(r)\right]\frac{r}{6} P_1(\mu) -
f' \frac{r}{5} \mu_{2, \alpha}(r) P_3(\mu) \notag
\end{align}
where a prime denote the derivative with respect to the comoving distance $\chi$, $P_\ell(\mu)$ is the Legendre polynomial of degree $\ell$, while the coefficients $\mu_\ell$ and $\mu_{\ell, \alpha}$ are
\begin{align}
\mu_{\ell}(r) &= \int \frac{k^2 dk}{2\pi^2} P_\text{m}(k) j_{\ell} (k\,r) G_\text{p}(k, L), \\
\mu_{\ell, \alpha}(r) &= \int \frac{k^2 dk}{2\pi^2} P_\text{m}(k) j_{\ell} (k\,r) G_\text{p}(k, L) (b_\alpha(k) - \bar{b}_\alpha),
\end{align}
and $G_\text{p}(k, L)$ is the Fourier transform of a spherical top-hat filter
modelling an isotropic binning of size $L = 4 h^{-1}$Mpc.

The Doppler correction is suppressed by a factor
$\mathcal{H}/k$ with respect to the terms discussed above.  Therefore, 
we can evaluate it safely in the flat-sky approximation
\begin{align}
\xi_\text{Doppler} &= \Biggl[(b_\text{Q} C_\alpha - \bar{b}_\alpha
C_\text{Q}) f - \frac{3}{5} (C_\text{Q} b_\text{v} - C_\alpha) f^2
\Biggr] \nu_1(r) P_1(\mu) + \\
& + \frac{2}{5}(C_\text{Q} b_\text{v} - C_\alpha) f^2  \nu_3(r) P_3(\mu) 
 - f C_\text{Q} \nu_{1, \alpha}(r) P_1(\mu), \notag
\end{align}
where $\nu_{\ell}$ and $\nu_{\ell, \alpha}$ are defined through
\begin{align}
\nu_{\ell}(r) &= \int \frac{k^2 dk}{2\pi^2}\left(\frac{\mathcal{H}}{k}\right) P_\text{m}(k) j_{\ell} (k\,r) G_\text{p}(k, L), \\
\nu_{\ell, \alpha}(r) &= \int \frac{k^2 dk}{2\pi^2}\left(\frac{\mathcal{H}}{k}\right) P_\text{m}(k) j_{\ell} (k\,r) G_\text{p}(k, L) (b_\alpha(k) - \bar{b}_\alpha), 
\end{align}
and the coefficients $C_\text{Q}$ and $C_\alpha$ are
\begin{align}
C_\text{Q} &= \Biggl( \frac{\mathcal{H'}}{\mathcal{H}^2} + \frac{2-5 s}{r\mathcal{H}} + 5 s - f^{\text{Q}}_{\text{evo}} \Biggr), \\
C_\alpha  &=   \Biggl(\frac{\mathcal{H'}}{\mathcal{H}^2} + 3 - f^{\alpha}_{\text{evo}} \Biggr) b_\text{R}. 
\end{align}

The evolution effect is computed assuming the following redshift-dependence for the biases of the tracers:
\begin{align}
\bar{b}_\alpha(z) &= \bar{b}_\alpha (2.4) \left(\frac{1+z}{1+2.4}\right)^{2.9}, \\
b_\text{Q}(z) &= 0.53 + 0.289(1+z)^2.
\end{align}
The redshift dependence of the Ly$\alpha$ forest has been modelled through observations~\cite{Busca2013, Slosar:2013fi, Slosar2011, McDonald:2004eu}, while the bias of the quasars 
follows the semi-empirical behaviour derived in Refs.~\cite{Croom:2004ui, Myers:2006dw}.
Note that the scale dependence of the Ly$\alpha$ bias is assumed to be redshift-independent. 

The Ly$\alpha$ bias at the mean redshift has been estimated from the best-fit parameters in Ref. \cite{Bourboux:2017cbm}, table 4 (auto-alone) and and its value is $\bar{b}_\alpha (2.4) = -0.135426$.
The quasar magnification and evolution biases
has been estimated in Ref. \cite{Irsic:2015} from
a fit for the quasars luminosity function used in
BOSS DR9 analysis \cite{Ross_2013}. 
Their values are set to $s = 0.295319$ and
$f^{\text{Q}}_{\text{evo}} = 5.7999$.

The evolution bias for the Ly$\alpha$ forest can be estimated analytically \cite{Irsic:2015}. The result
gives $f^{\alpha}_{\text{evo}} \simeq -3$.

\begin{figure}
\centering
    \begin{subfigure}[b]{0.485\textwidth}
\includegraphics[width=\textwidth]{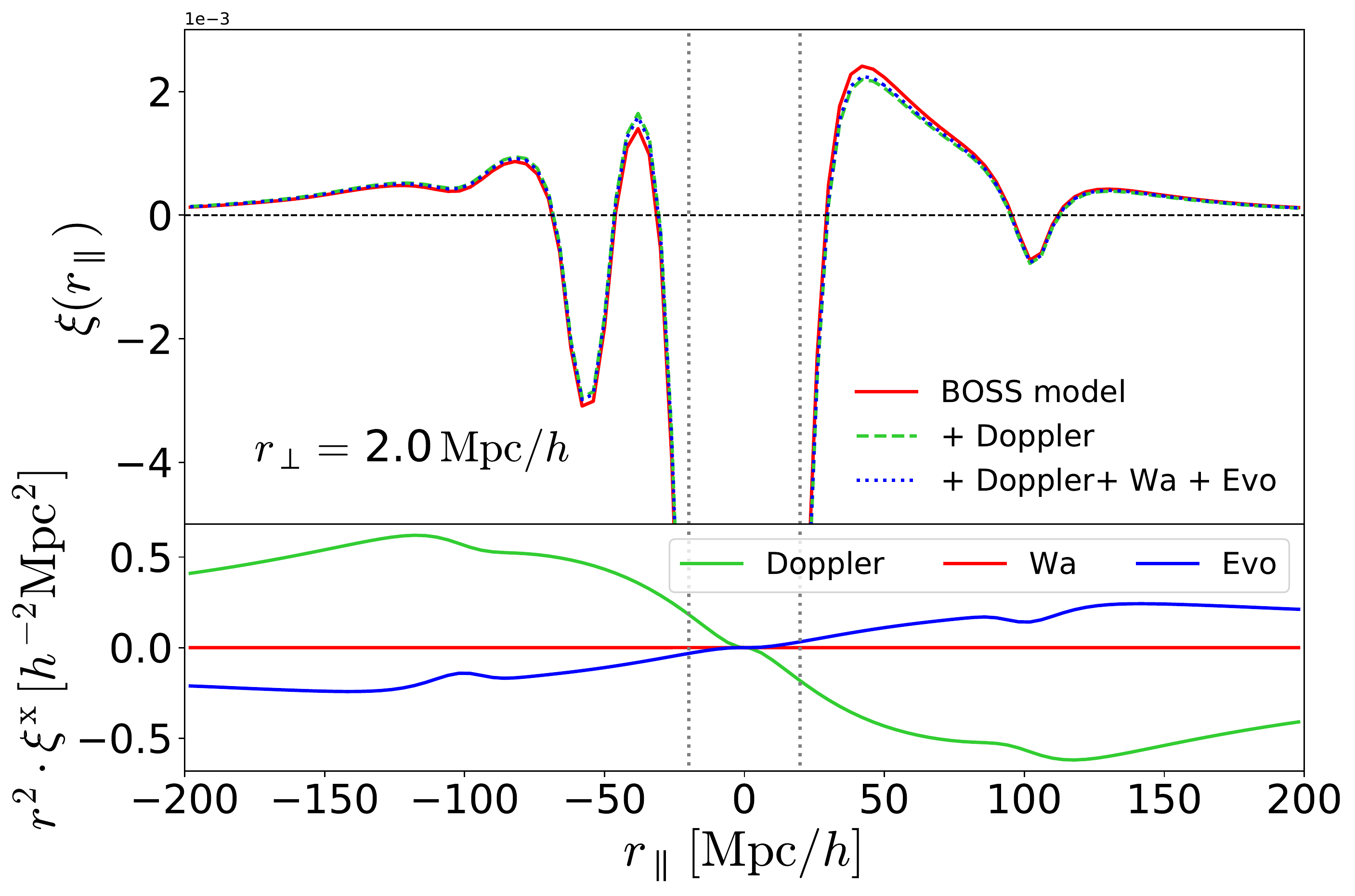}
        \caption{$r_\perp = 2 \,\text{Mpc}/h$}
        \label{fig:rel-rt2}
    \end{subfigure}
    \begin{subfigure}[b]{0.5\textwidth}
\includegraphics[width=\textwidth]{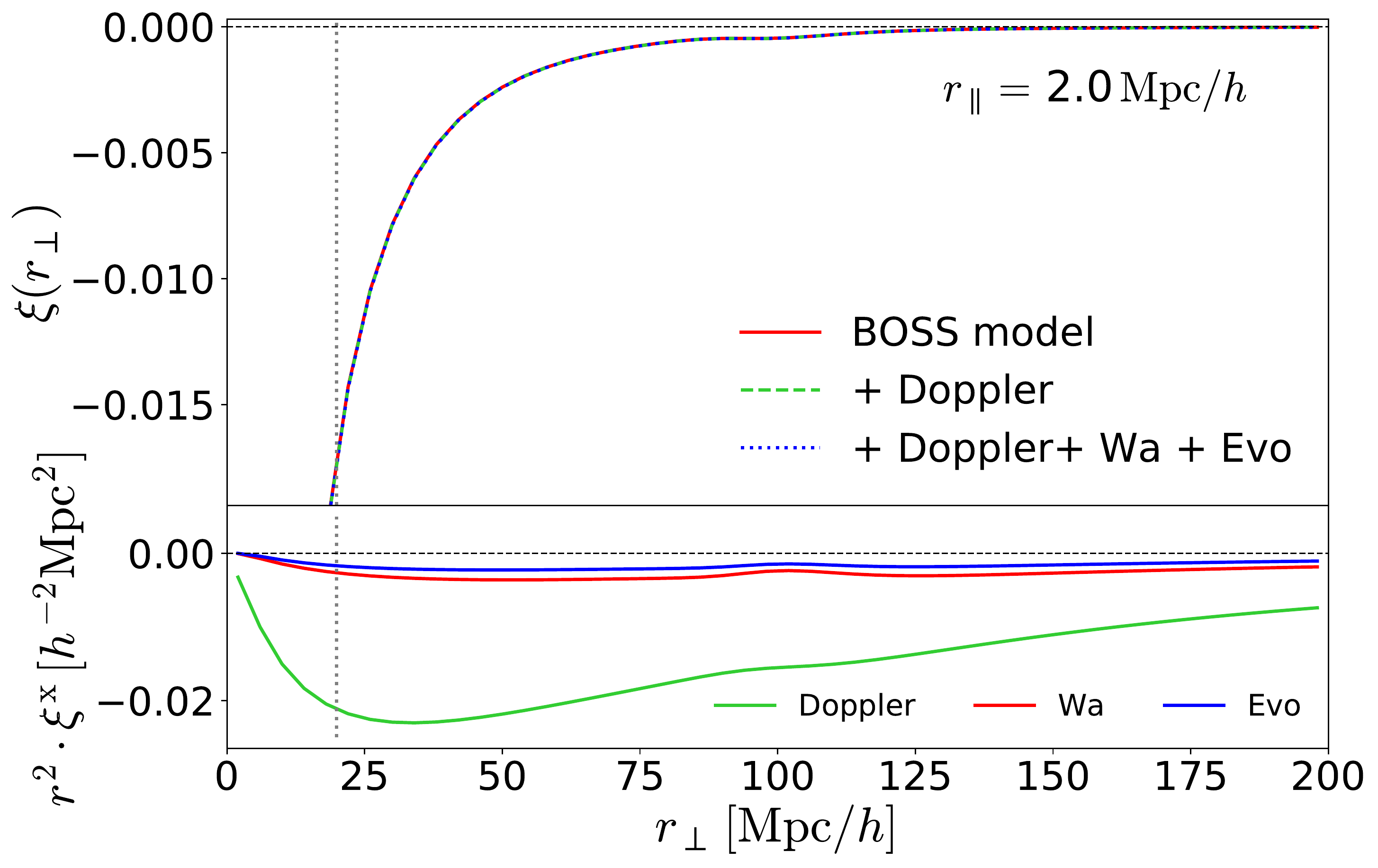}
        \caption{$r_\parallel = 2\, \text{Mpc}/h$}
        \label{fig:rel-rp2}
    \end{subfigure}
    \caption{Top panels: QSOs-Ly$\alpha$ cross-correlation
for $r_\perp = 2\,\mathrm{Mpc}/h$ (left panel) and $r_\parallel = 2\,\mathrm{Mpc}/h$ (right panel) as modelled by BOSS (red line), the BOSS model with Doppler correction (green dashed lines) and with Doppler, wide-angle and evolution corrections (blue dotted lines).  
The distortion matrix has not been applied to 
the correlation function. 
Bottom panels: Doppler (green), wide-angle (red) and bias evolution effect (blue) for the two configurations. 
The mean redshift has been fixed to be $z_\text{mean} = 2.4$. Vertical dotted lines represent a typical cut scale $r_\mathrm{cut} = \sqrt{r^2_{\parallel, \text{cut}}  + r^2_{\perp, \text{cut}} } = 20\,\mathrm{Mpc}/h$.
}
\label{fig:RelEff}
\end{figure}

In Fig.~\ref{fig:RelEff} we show the Doppler, wide-angle and bias evolution corrections to
the Ly$\alpha$-QSOs cross-correlation as modelled 
by BOSS. The left panel shows the correlation function
in terms of the longitudinal separation $r_\parallel$, for
$r_\perp = 2 \,\text{Mpc}/h$, while in the right 
panel we show the transverse correlation function, 
at fixed longitudinal separation $r_\parallel = 2 \,\text{Mpc}/h$. 
Wide-angle and z-evolution correction have little impact compared to the Doppler effect for both 
configurations. In fact, these corrections are suppressed by a factor $r/\chi_\text{m}$
and we do not expect them to have a large impact at
high redshift. 
In particular, we observe that in the configuration $r_\perp = 2 \,\text{Mpc}/h$ wide-angle effects are totally negligible, as expected being an almost longitudinal configuration.

\section{Signal-to-noise analysis}
\label{sec:sn}
In this section we compute the cumulative signal-to-noise of the Doppler signal, the wide-angle signal and the evolution correction
for BOSS, eBOSS and DESI. 
The cumulative signal-to-noise is given by
\begin{equation}
\left(\frac{S}{N} \right)^2 =  \xi^i_\text{X} \mbox{C}^{-1}_{ij} \xi^j_\text{X}\, ,  \label{SN}
\end{equation}
where $\xi_\text{X}$ is the signal we are seeking and the 
indices $i,j$ run over the pixels with separation $r > r_\text{min}$. We will consider the three contributions to the correlation function discussed in the previous section:
the Doppler relativistic effect, the wide-angle and the redshift evolution corrections.
We will show the cumulative signal-to-noise 
as a function of the minimum scale $r_\text{min}$ that we
are able to measure and model for a given survey. 

The covariance matrix $\mbox{C}$ 
depends on the specifics of the survey. For BOSS we use
the expression in Ref.~\cite{Bourboux:2017cbm}, Eq.~17. This expression provides a good fit for its
diagonal elements.

In order to estimate the covariance for eBOSS and DESI, we assume that the diagonal elements
scale as follows:
\begin{equation}
\text{C}_{\text{eBOSS}/\text{DESI}}= \text{C}_\text{BOSS} \left (\frac{N_\text{Q} N_\alpha}{V^2} \right)_\text{BOSS} \left (\frac{N_\text{Q} N_\alpha}{V^2} \right)^{-1}_{\text{eBOSS}/\text{DESI}},
\label{cov-rescale}
\end{equation}
where $N_\text{Q}$ and $N_\alpha$ are the number 
of quasars and forests, respectively, while $V$
denotes the volume of the survey.
The number of quasars and forests are 
computed from Ref.~\cite{Blomqvist:2019rah} and Ref.~\cite{Aghamousa:2016zmz} for
eBOSS and DESI, respectively.
A summary of the specifics used in our analysis
is reported in table \ref{table_surveys}.

\begin{table}[tbp]
\centering
\begin{tabular}{|c|c|c|c|c|}
\hline
\multicolumn{1}{|c|}{}                    & 
 \multicolumn{1}{c|}{$f_\text{sky}$}             &
  \multicolumn{1}{c|}{$z$-range}            &
\multicolumn{1}{|c|}{$N_\text{Q}$}                    & 
 \multicolumn{1}{c|}{$N_\alpha$}      \\
\hline
BOSS  & $0.24$  & $[1.85, 3.5]$ & $217780$ & $157845$ \\
eBOSS & $0.24$  & $[2.05, 3.5]$ & $270816$ & $257245$ \\
DESI  & $0.34$ & $[1.96, 3.55]$ & $787227$ & 700000
\\
\hline
\end{tabular}
\caption{
Specifics for BOSS, eBOSS and DESI used to 
estimate the signal-to-noise in Fig. \ref{fig:SNplot}. 
}
\label{table_surveys}
\end{table}

The expression in \eqref{cov-rescale} is a good approximation because the Ly$\alpha$-QSOs covariance is shot-noise dominated.
However, the non-diagonal elements in 
the covariance will slightly suppress
the signal-to-noise ratio. We correct for this effect by assuming that the impact of the non-diagonal elements of
the covariance in the signal-to-noise ratio for the Doppler effect
is the same for the 3D correlation function and for the dipole of the correlation function. This assumption is well justified by considering that the Doppler term is the leading contribution of the odd multipoles of the power spectrum or 2-point correlation function. Between the odd multipoles, the dipole is carrying most of the signal-to-noise. In Appendix \ref{apA} we computed the S/N for the dipole with the full covariance and for the diagonal covariance. We found that the non-diagonal elements suppress the S/N
up to a factor 5/7. Therefore, we apply this suppression factor to our S/N. 
It is worth remarking that this is a conservative choice, in particular on large scale where covariance matrix tends to be more diagonal.

\begin{figure}
\centering
\includegraphics[width=0.7\textwidth]{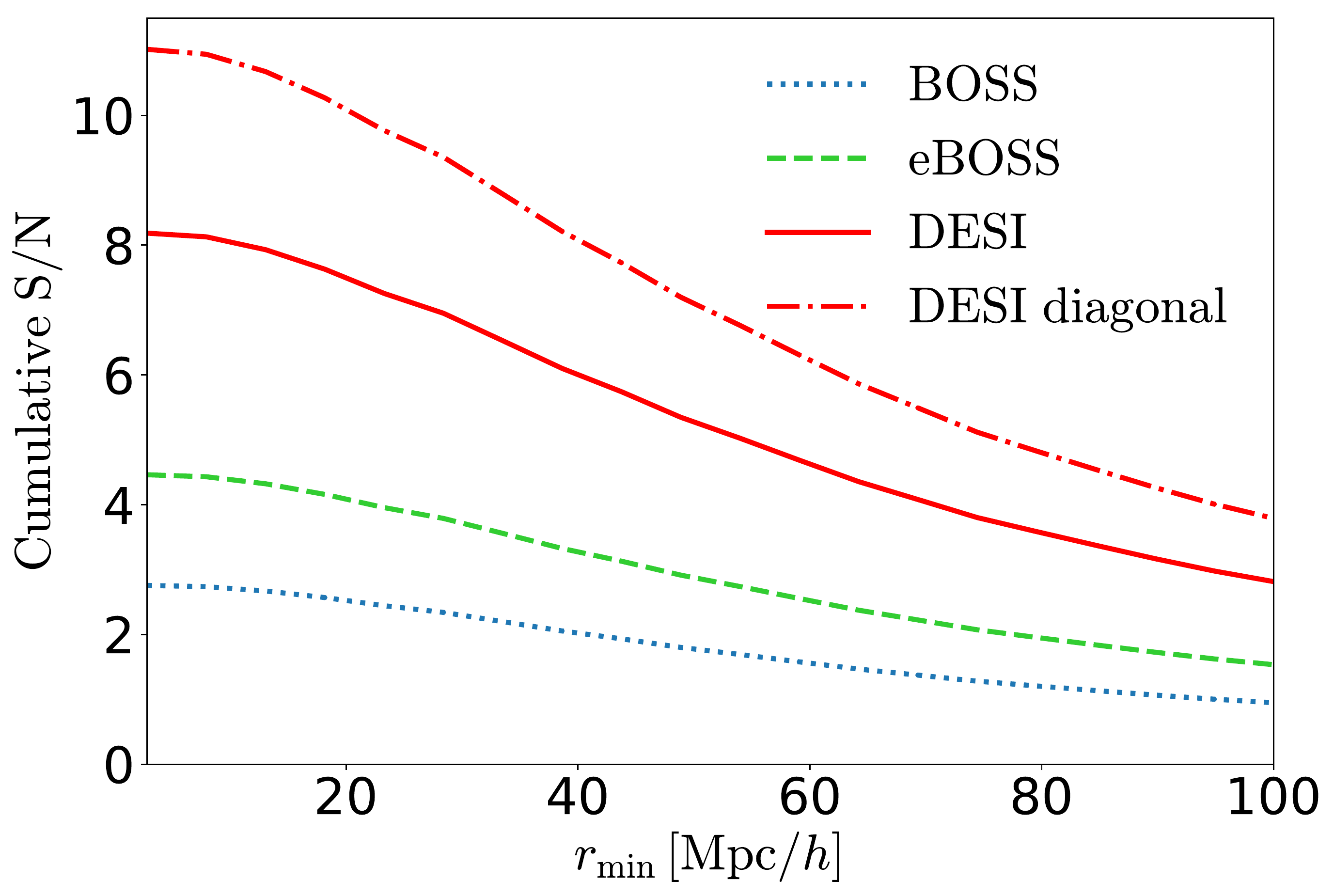}
\caption{Cumulative signal-to-noise as a function of the minimum distance included in the analysis.}
\label{fig:SNplot}
\end{figure}

In Fig.~\ref{fig:SNplot} we show the 
total cumulative signal-to-noise for
three different surveys: BOSS, eBOSS and DESI.
We find that the cumulative signal-to-noise for BOSS
amounts to approximately 2 for scales  $r_{\min} > 20 \ {\rm Mpc}/h$. Therefore, we do not expect the signal to be detectable from BOSS catalogue, in agreement with the lack of relativistic detection so far.

For eBOSS we expect a signal-to-noise ratio close
to 4. In the first eBOSS analysis of the Ly$\alpha$-QSOs cross-correlation  (see Ref.~\cite{Blomqvist:2019rah}) a dipole contribution is included in the fit for the correlation function.
The dipole models the relativistic effects discussed in our work. The best-fit support a non-zero contribution of the relativistic dipole. However, correlations with other systematic 
effects prevent a significant
detection with current data. 
We remark that, differently from the eBOSS collaboration, we have not marginalized over the Alcock Paczy\'nski parameters. Therefore our signal-to-noise ratio could be slightly more optimistic than the analysis performed in Ref.~\cite{Blomqvist:2019rah}.

The signal-to-noise ratio for DESI is more promising. A significant improvement in 
the number of quasars and forests leads to
an expected $S/N \approx 7$. 
We show for comparison also the signal-to-noise for
DESI when we consider a diagonal covariance. Our results indicate that surveys like DESI and WEAVE-QSO will need to account for some relativistic effects in their data analysis to accurately model the Lyman-α QSOs cross correlation.

From Fig.~\ref{fig:SNplot} we clearly see that the signal-to-noise ratio is limited by the shot-noise on small scales. The Doppler term is suppressed by a factor $\HH/k$ with respect to the matter power spectrum and, therefore, weights more large scale power. As a consequence, at small scales the relativistic signal-to-noise ratio is more strongly affected by shot-noise. On the other hand, more conservative non-linear scale cuts are not heavily disadvantaged.

Compared to the Fourier space analysis in Ref.~\cite{Irsic:2015}, we used a more conservative
estimation of the Ly$\alpha$ noise and a more realistic estimation for the covariance matrix\footnote{We warn the reader that even if we perform a more realistic and conservative analysis we obtain roughly the same signal-to-noise ratio of Ref.~\cite{Irsic:2015} due to slightly different number of sources. In appendix~\ref{apA} we consider the same specifications of Ref.~\cite{Irsic:2015}.}.

\begin{figure}
    \begin{subfigure}[b]{0.49\textwidth}
\includegraphics[width=\textwidth]{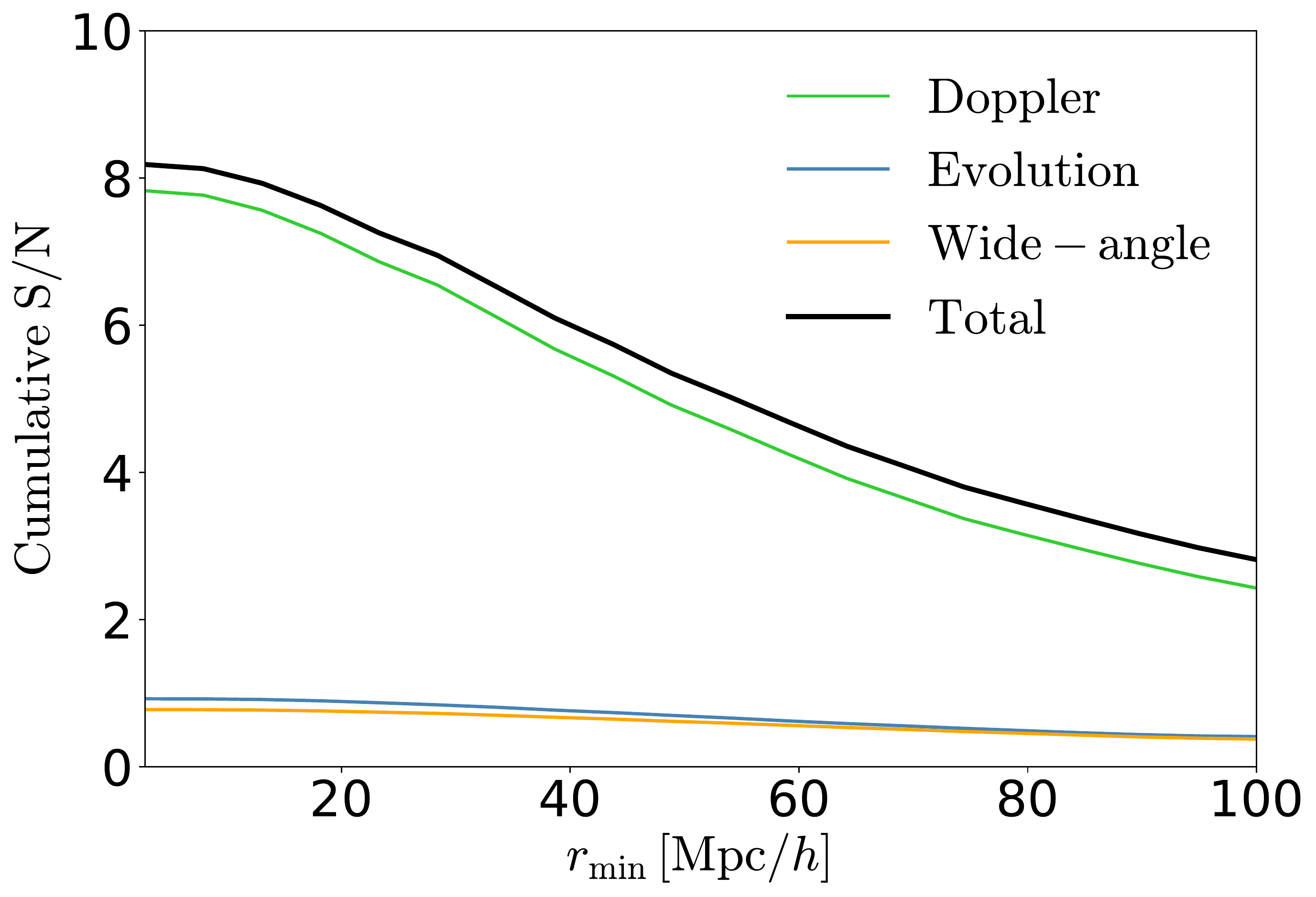}
        \caption{}
        \label{fig:sn2-a}
    \end{subfigure}
    \begin{subfigure}[b]{0.49\textwidth}
\includegraphics[width=\textwidth]{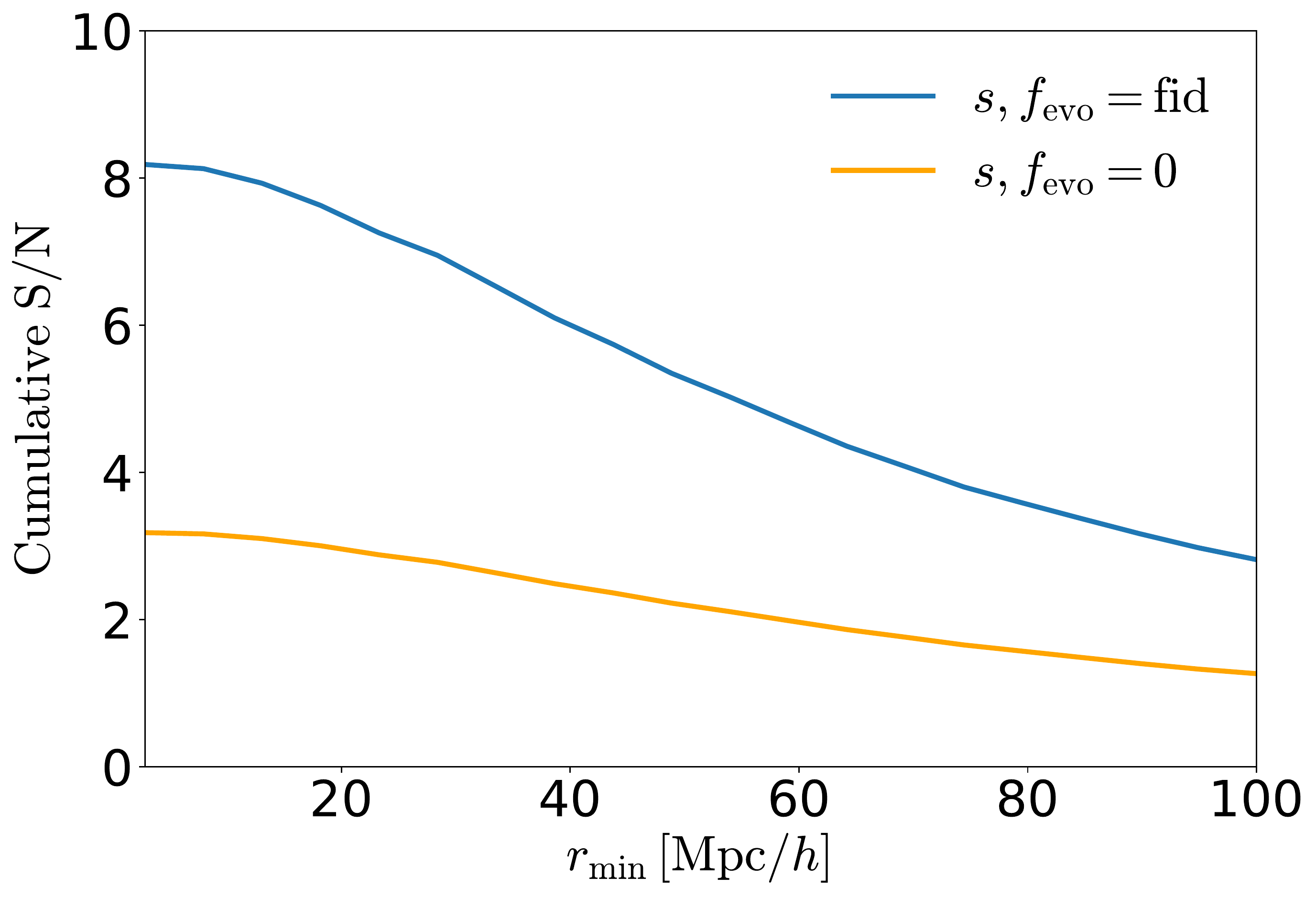}
        \caption{}
        \label{fig:sn2-b}
    \end{subfigure}
    \caption{Left panel: contributions of the different terms to the cumulative signal-to-noise of the relativistic effect in Ly$\alpha$-QSOs cross-correlation, we include the Doppler relativistic effect, the wide-angle and redshift-evolution correction. Right panel: expected Doppler signal-to-noise for different values of the evolution and magnification biases, the blue line is computed assuming fiducial values for the biases, the orange line is obtained setting them to zero and the signal-to-noise has been computed assuming DESI specifications.}
\label{fig:SNplot2}
\end{figure}

In Fig.~\ref{fig:SNplot2} (left panel) we compare the different contributions to the signal-to-noise. 
Both wide-angle and evolution effects 
have very little impact on the total signal-to-noise, which is dominated by the Doppler
effect. 
The right panel in Fig.~\ref{fig:SNplot2} 
shows the signal-to-noise for DESI, computed by
setting to zero the quasars magnification bias and the evolution bias of the two tracers and we
compare it to the $S/N$ obtained for the fiducial values of the biases. 
We observe that the signal-to-noise is suppressed from $S/N \sim 7$ to $S/N \sim 2$, when magnification and evolution biases are set to zero. This
shows that it will be crucial to properly model astrophysical biases if we want extract physical 
information from the detection of relativistic effects. 

\section{Shift in the best-fit parameters}
\label{sec:fisher}
In this section we will study the impact of neglecting the Doppler corrections in the theoretical model for the
cross-correlation. 
Our approach is based on
 the Fisher matrix information formalism~\cite{Fisher:1935bi, Tegmark:1997yq}. 
The Fisher information matrix is defined as
\be
\text{F}_{ij} = -\mean{\frac{\partial^2 \mathcal{L}}{\partial p_i \partial p_j}},
\ee
where $\mathcal{L}(\mathbf{x}, \mathbf{p})$ is the 
probability distribution of our measurement $\mathbf{x}$, and
$\mathbf{p}$ is a set of parameters that we aim to
measure. The inverse of the Fisher matrix
is the most optimistic covariance matrix
for the set of parameters $\mathbf{p}$.

We apply the Fisher formalism to an experiment
that measures the Ly$\alpha$-QSOs cross-correlation function, like DESI and BOSS. 
In this application, we can express the Fisher matrix as
\be
\text{F}_{ij} = \frac{\partial \xi^a}{\partial p_i} \text{C}^{-1}_{ab} \frac{\partial \xi^b}{\partial p_j},
\ee
where $\xi$ is the model for the correlation function 
and $\text{C}$ is the covariance for our measurements. 
Neglecting Doppler effects in the model
for the cross-correlation will result in a biased parameter estimation by shifting the best-fit parameters.
In the Fisher matrix formalism, this shift can be 
expressed~\cite{Taylor:2006aw} as 
\be
\Delta p_i = \text{F}^{-1}_{ij} \frac{\partial \xi^a}{\partial p_j}  \text{C}^{-1}_{ab} \Delta \xi^b,
\ee
where $\Delta \xi$ is the contribution of Doppler
effects to the correlation function. 

In our analysis, we
fix the cosmological parameters to be the fiducial
values in Ref.~\cite{Bourboux:2017cbm}, and we consider the set of parameters that are fitted in the model for the correlation function in Ref.~\cite{Bourboux:2017cbm}.
These set of parameters, summarised in Ref.~\cite{Bourboux:2017cbm}, table 4, includes
the Ly$\alpha$ clustering bias parameters $(\bar{b}_\alpha, b_\Gamma)$, the RSD parameter $\beta_\alpha$, all the parameters describing 
the metals, HCD, proximity-effect contributions and
a systematic effect in the estimation of the 
quasar redshift $\Delta r_\parallel$. 

The fiducial values for our set of parameters have been fixed to be the best-fit estimated from the auto-correlation alone. Quasars parameters has been fixed to the values $b_Q = 3.87$ and $\sigma_\text{v} = 6.43$. The systematic error on the quasars redshift
has been set to be zero in the fiducial model.

\begin{figure}
\centering
\includegraphics[width=\textwidth]{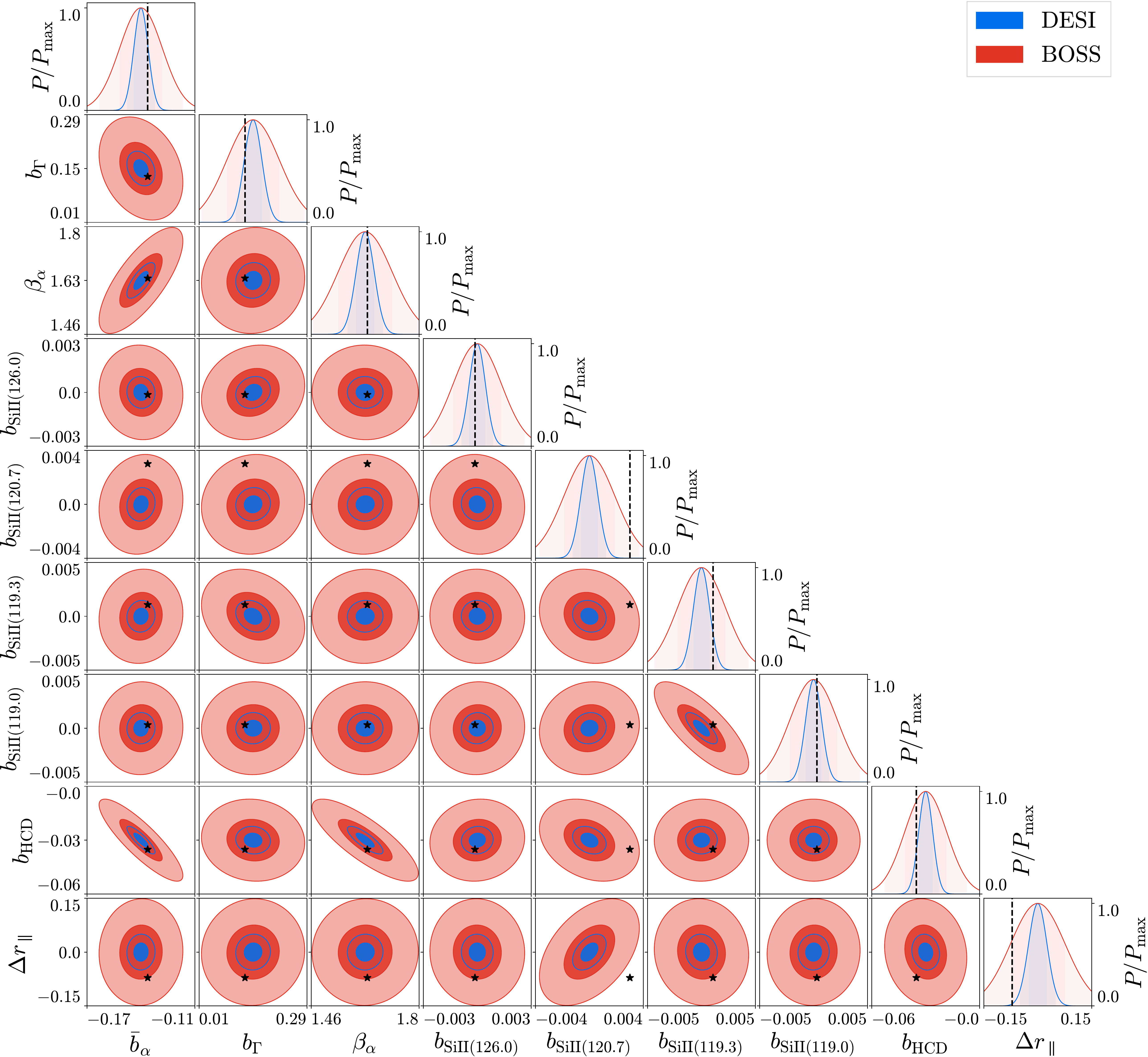}
\caption{1-$\sigma$ and 2-$\sigma$ contour plot
estimated for BOSS and DESI with Fisher matrix analysis for a subset of the parameters included in the analysis. A black star denotes the shift in the best-fit parameters that we expect when Doppler effects are not included in the model for a DESI-like survey.
The axis ticks refer to the range limits in the 2D contour plots. 
This plot is produced with the python library {'CosmicFish'} \cite{Raveri:2016xof, Raveri:2016leq}.
}
\label{fig:fisher}
\end{figure}

In Fig. \ref{fig:fisher} we show the contour plot for 
DESI (blue contours) and BOSS (red contours).
The black star highlight the shift in the best-fit 
parameters that we expect for a survey like DESI. 
The parameters that are more degenerate with the Doppler
correction are the systematic error on the estimation of the QSOs redshift, whose
best-fit is biased approximately of 3-$\sigma$, and 
the clustering bias of one of the metals (SiII (120.7)).
The degeneracy with $\Delta r_\parallel$ was also pointed out in Ref.
\cite{Blomqvist:2019rah}, and it was the main uncertainty which prevented the detection of 
a relativistic dipole in the cross-correlation
for eBOSS. 
In general, we observe that neglecting the Doppler contribution to the Lyman-$\alpha$ QSOs cross-correlation can contaminate by a non-negligible amount the parameter estimation for several bias parameters in a DESI-like survey.
\clearpage
\section{Conclusions}
\label{sec:conc}
In this paper we forecast the relevance of relativistic effects in the cross-correlation of the Lyman-$\alpha$ forest and the quasars number counts for DESI. 

Our work extends and improves the analysis presented in Ref.~\cite{Irsic:2015}, where the signal-to-noise for the imaginary part of the Fourier space spectrum has been
computed for BOSS and DESI surveys, neglecting any systematic effects that we have introduced in the current work.

In agreement with Ref.~\cite{Irsic:2015}, we found that the major contribution to relativistic correction
is sourced by the Doppler effect, while wide-angle and bias-evolution effects are significantly smaller. 
This is due to the fact that at high redshift 
these corrections are suppressed by a factor $r/\chi_\text{m}$, where $r$ is the separation between the tracers and $\chi_\text{m}$ is the comoving distance to the effective redshift of the sample.  

For a survey like DESI, we forecast a signal-to-noise
for the relativistic corrections to be $S/N \sim 7$,
provided the fact that we will be able to model accurately several astrophysical parameters that 
source our signal (especially the evolution biases of the two tracers).

Finally, we studied the impact of relativistic effects
on parameters estimation. Since we expect that the
relativistic effects do not significantly shift
the position of the BAO peak, we focused on the 
bias parameters and astrophysical systematics in the cross-correlation. 
In fact, some of the
parameters modelling systematics show 
a mild tension between the best-fit values estimated from the cross-correlation QSOs-Ly$\alpha$ and the Ly$\alpha$
auto-correlation, where the relativistic effects here considered can be safely neglected \cite{Bourboux:2017cbm}.  
We have shown that some bias parameters, in particular the quasar redshift estimation parameter $\Delta r_\parallel$, measured through the Lyman-$\alpha$ QSOs correlation are sensitive to the Doppler effect. 
However, we find that the tensions between auto-correlation and cross-correlation analysis described above cannot be resolved including the Doppler effect in
the theoretical model for the cross-correlation. 
Our results indicate that surveys like DESI and WEAVE-QSO will need to account for some relativistic effects in their data analysis to accurately model the Ly$\alpha$-QSOs cross correlation.

\acknowledgments{We thank Jim Rich for useful discussions, and for having provided us with the BOSS distortion matrix information. 
We thank the anonymous referee for reading the paper carefully and providing thoughtful comments.
FL and ED (No.~171494 and~171506) acknowledges financial support from the Swiss National Science Foundation. MV is supported by INFN INDARK PD51 grant and acknowledges financial contribution from the agreement ASI-INAF n.2017-14-H.0. VI acknowledges support by the Kavli Foundation, and thanks US NSF grant AST-1514734.
}

\vspace{1.5cm}

\appendix

\noindent{\LARGE \bf Appendix}

\section{Lensing dipole}
\label{ap-lens}
The lensing contribution to the quasars' number count introduces an asymmetric contribution
to the cross-correlation Ly$\alpha$-QSOs: the background quasars will be lensed 
by the neutral hydrogen clouds where a Ly$\alpha$ transition occurs, while the Ly$\alpha$ flux fluctuations are not lensed, to linear order, by the foreground quasars.
The lensing contribution in the cross-correlation of two galaxy populations was computed in \cite{Bonvin:2013} and can be adapted to the Ly$\alpha$-QSOs cross-correlation:
\begin{equation}
\xi^\text{lens}(z, r, \beta) = (1+z) \frac{3\Omega_m \pi}{4}
b_\alpha (5s - 2) \, r \, \mathcal{H}_0\, \cos{\beta} \,\Theta\left(D_\text{Q}-D_\alpha\right) \mu_\text{lens}(\beta), \label{lens_xi}
\end{equation}
where $\Theta$ is the Heaviside function and the function $\mu_\text{lens}$ is
\begin{equation}
\mu_\text{lens}(\beta) =  \int_{0}^{\infty} k_\perp d k_\perp\HH_0 P(k_\perp) J_0\left(k_\perp\, r \,\sin{\beta}\right),
\end{equation}
being $J_0$ the order-0 Bessel function. 
Eq.~\eqref{lens_xi} assumes the Limber approximation and it is the lowest order contribution in $r/\chi_\text{m}$. 
In order to estimate the lensing contribution
to the anti-symmetric part of the correlation function, we compare the Doppler and the lensing dipole. 
\begin{figure}[h!]
\centering
\includegraphics[width=0.7\textwidth]{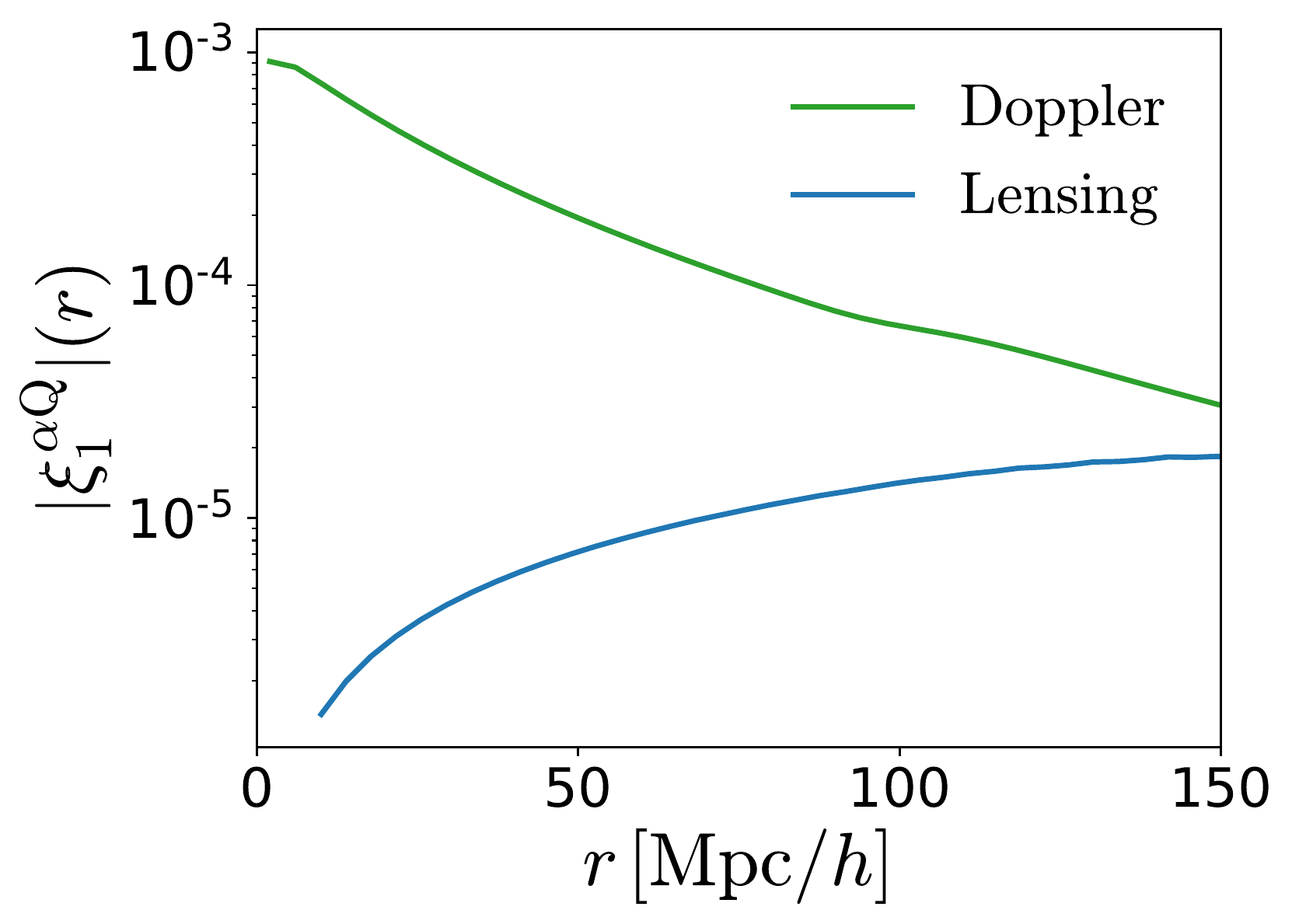}
\caption{Comparison between Doppler dipole and lensing dipole at $z = 2.4$. 
}
\label{fig:lens-dip}
\end{figure}
In Fig.~\ref{fig:lens-dip} we show the Doppler and lensing dipole for the Ly$\alpha$-QSOs cross-correlation.
Despite the fact that lensing is expected to be important for tracers at high redshift, the Doppler contribution dominates over the lensing up to $r \sim 150\,\text{Mpc}/h$. 
This is due to the bias factor $b_\alpha (5s - 2) \sim 6 \times 10^{-2}$, which suppresses the lensing signal. 
Therefore, we can safely neglect lensing in our signal-to-noise analysis.

\section{Signal-to-noise for the dipole}
\label{apA}
In this section we compare the real space signal-to-noise analysis of the dipole to the Fourier space analysis presented in Ref.~\cite{Irsic:2015}. 
We assume the same specifics
as Ref.~\cite{Irsic:2015} to model
BOSS and DESI covariance. 

The cumulative signal-to-noise for the dipole
can be computed as described in Sec.~\ref{sec:sn}, while the dipole covariance has been computed
from the general expression in Ref.~\cite{Hall:2016}.

\begin{figure}[h!]
\centering
    \begin{subfigure}[b]{0.5\textwidth}
        \includegraphics[width=\textwidth]{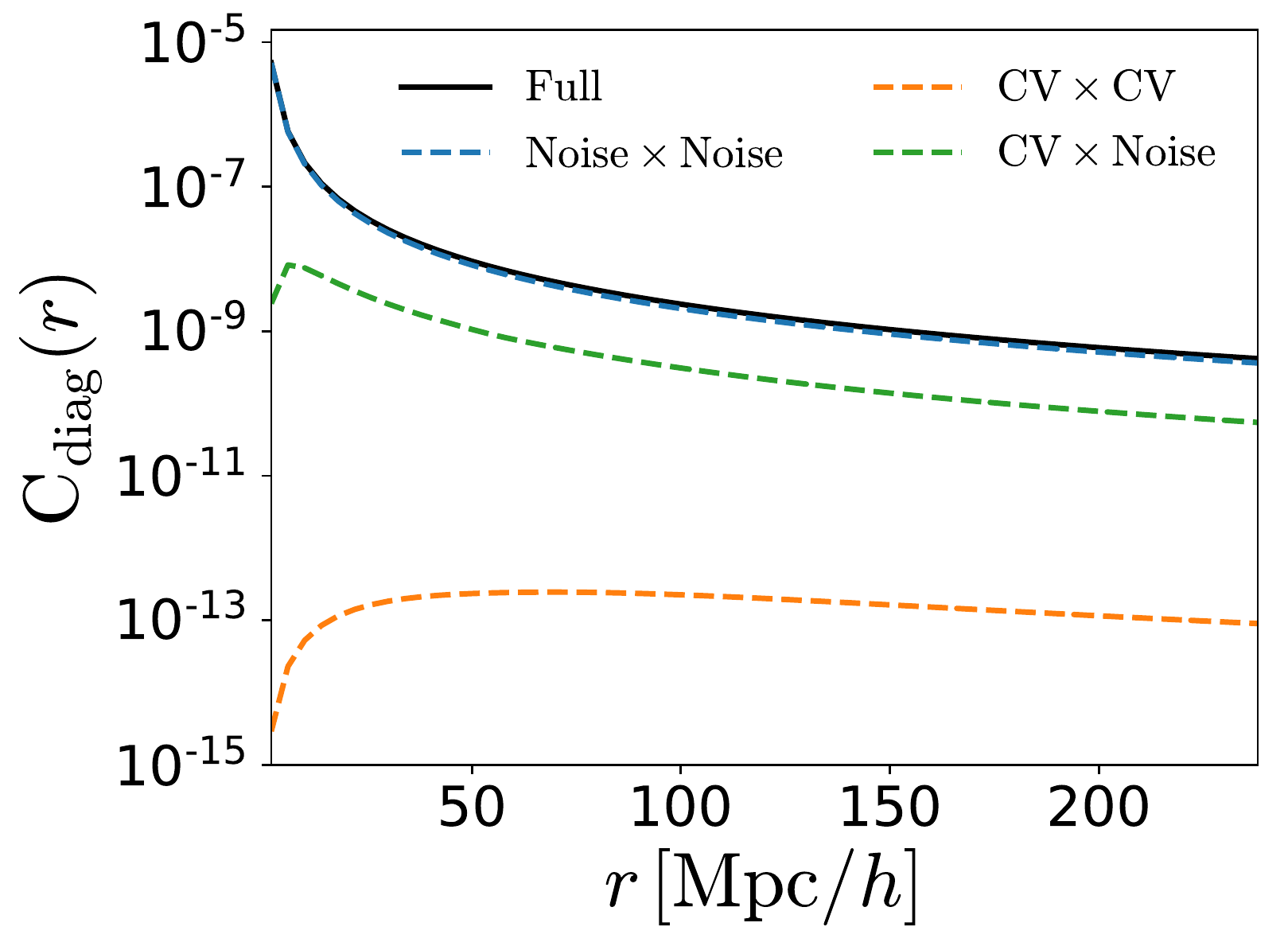}
    \end{subfigure}
    \begin{subfigure}[b]{0.47\textwidth}
        \includegraphics[width=\textwidth]{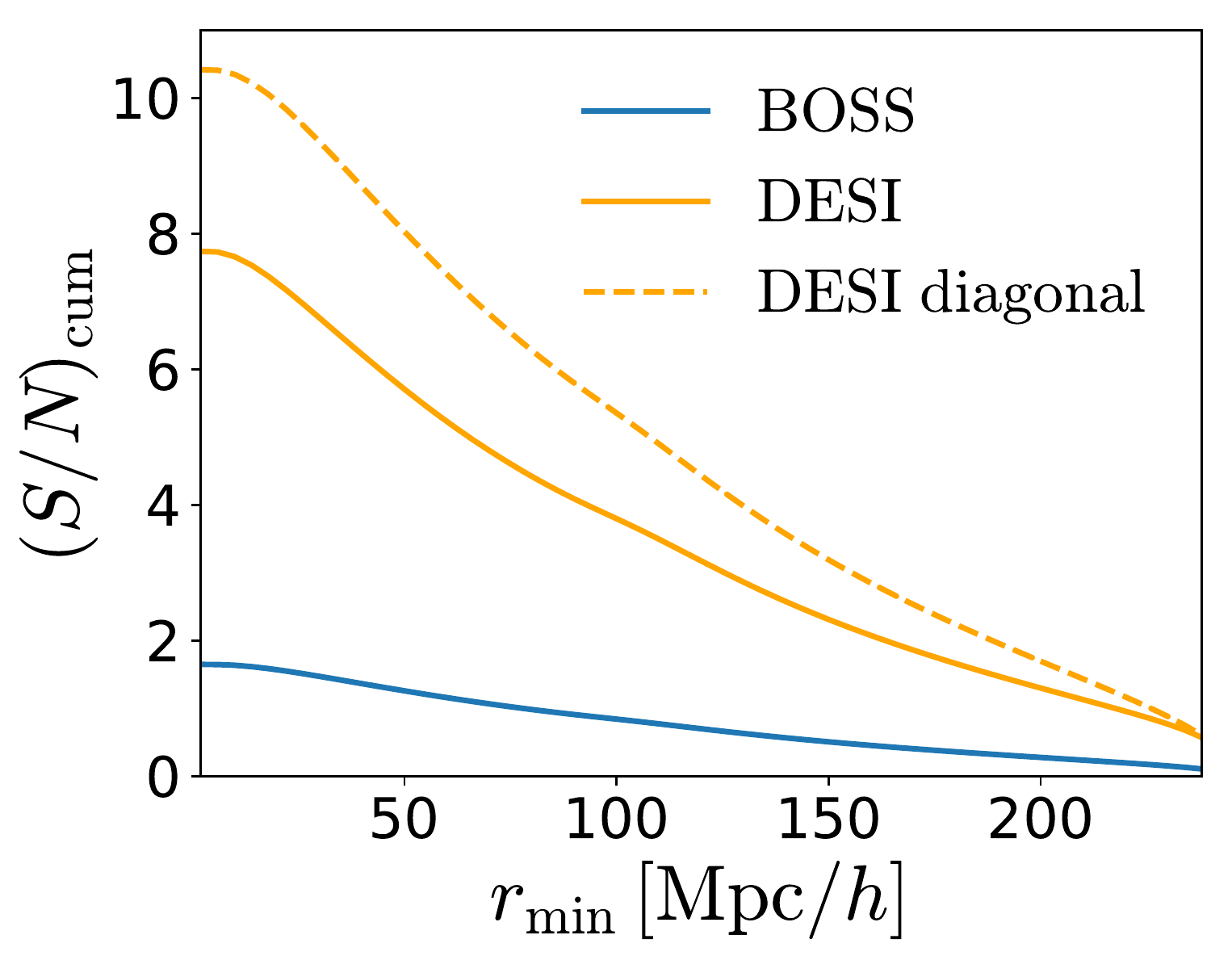}
    \end{subfigure}
    \caption{Left panel: diagonal elements of the covariance for the dipole. Different colors denote different contributions (cosmic variance $\times$ cosmic variance - orange; cosmic variance $\times$ shot-noise - green;
    shot-noise $\times$ shot-noise - blue; total - black). Right panel: cumulative signal-to-noise ratio for BOSS (blue) and DESI (orange).  The specifics for the two surveys follow tha Fourier space analysis in Ref. \cite{Irsic:2015}. The dashed orange line represents the signal-to-noise for DESI, when only the diagonal elements of the covariance are taken into account. 
}
    \label{fig:dipole}
\end{figure}

In Fig.~\ref{fig:dipole} (left panel) we compare the different contribution to the covariance matrix. The diagonal elements are dominated by 
the purely shot-noise contribution. 
Therefore, we show that our approach to estimate 
the covariance for DESI in Sec.~\ref{sec:sn} is 
legitimate. 

The right panel in Fig.~\ref{fig:dipole} shows the
signal-to-noise of the dipole for BOSS (blue line)
and DESI (orange line). The signal-to-noise of the
dipole gives consistent results with the Fourier space analysis in Ref.~\cite{Irsic:2015}. 
The dashed orange line is the signal-to-noise
computed considering only the diagonal elements. 
We found that for the relativistic dipole the non-diagonal elements suppress the signal-to-noise up to a factor $5/7$.

\bibliographystyle{JHEP}
\bibliography{mybib}

\end{document}